\documentclass[dvipsnames,conference]{IEEEtran}

\usepackage[T1]{fontenc}
\usepackage[scaled=.83]{beramono}

\usepackage[colorinlistoftodos]{todonotes}
\usepackage[inference]{semantic}
\usepackage{halloweenmath}
\usepackage{fontawesome5}
\usepackage{listofitems}
\usepackage{breakcites}
\usepackage{glossaries}
\usepackage{hyperref}
\usepackage{cleveref}
\usepackage{stmaryrd}
\usepackage{marvosym}
\usepackage{listings}
\usepackage{amssymb}
\usepackage{amsmath}
\usepackage{nameref}
\usepackage{amsthm}
\usepackage{xspace}
\usepackage{xfrac}
\usepackage{tikz}
\usepackage{soul}
\usepackage{bm}
\usepackage{paralist}

\hypersetup{
    colorlinks,
    linkcolor={red!50!black},
    citecolor={blue!50!black},
    urlcolor={blue!80!black}
}

\setul{0.95ex}{0.3ex}

\usetikzlibrary{calc,decorations.pathmorphing,shapes,positioning}
\newcounter{sarrow}

\makeatletter
\newcommand{\xRightarrow}[2][]{\ext@arrow 0359\Rightarrowfill@{#1}{#2}}
\makeatother

\newcommand{\Thmref}[1]{\Cref{#1}~(\nameref{#1})}

\newenvironment{nscenter}
 {\parskip=0pt\par\nopagebreak\centering}
 {\par\noindent\ignorespacesafterend}

\newcommand{\pages}[1]{}

\newcommand{\irccol}[0]{Apricot}
\newcommand{\ulccol}[0]{RedOrange}
\newcommand{\objcol}[0]{Emerald} 

\newcommand{\irdcol}[0]{CarnationPink}

\newcommand{\col}[2]{\ensuremath{{\color{#1}{#2}}}}

\newcommand{\src}[1]{\ensuremath\mathsf{\col{\stlccol}{#1}}}
\newcommand{\irl}[1]{\ensuremath\mathit{\col{\irccol}{#1}}}
\newcommand{\trg}[1]{\ensuremath\mathbf{\col{\ulccol}{#1}}}
\newcommand{\obj}[1]{\ensuremath\mathtt{\col{\objcol}{#1}}}
\newcommand{\ird}[1]{\ensuremath\mathit{\col{\irdcol}{#1}}}

\newcommand{\mb}[1]{\ensuremath{\mathbb{#1}}}

\newcommand{\bul}[1]{{\setulcolor{RoyalBlue}\ul{#1}}}
\newcommand{\rul}[1]{{\setulcolor{RedOrange}\ul{#1}}}
\newcommand{\iul}[1]{{\setulcolor{Apricot}\ul{#1}}}
\newcommand{\oul}[1]{{\setulcolor{Emerald}\ul{#1}}}

\newcommand{\lock}{\ensuremath\text{\scriptsize\faIcon{lock}}}
\newcommand{\unlock}{\ensuremath\text{\scriptsize\faIcon{lock-open}}}

\newcommand{\isdef}[0]{\ensuremath{\mathrel{\overset{\makebox[0pt]{\mbox{\normalfont\tiny\sffamily def}}}{=}}}}

\newcounter{contrib}

\newcommand{\BareCoqSymbol}{\includegraphics[height=0.9em]{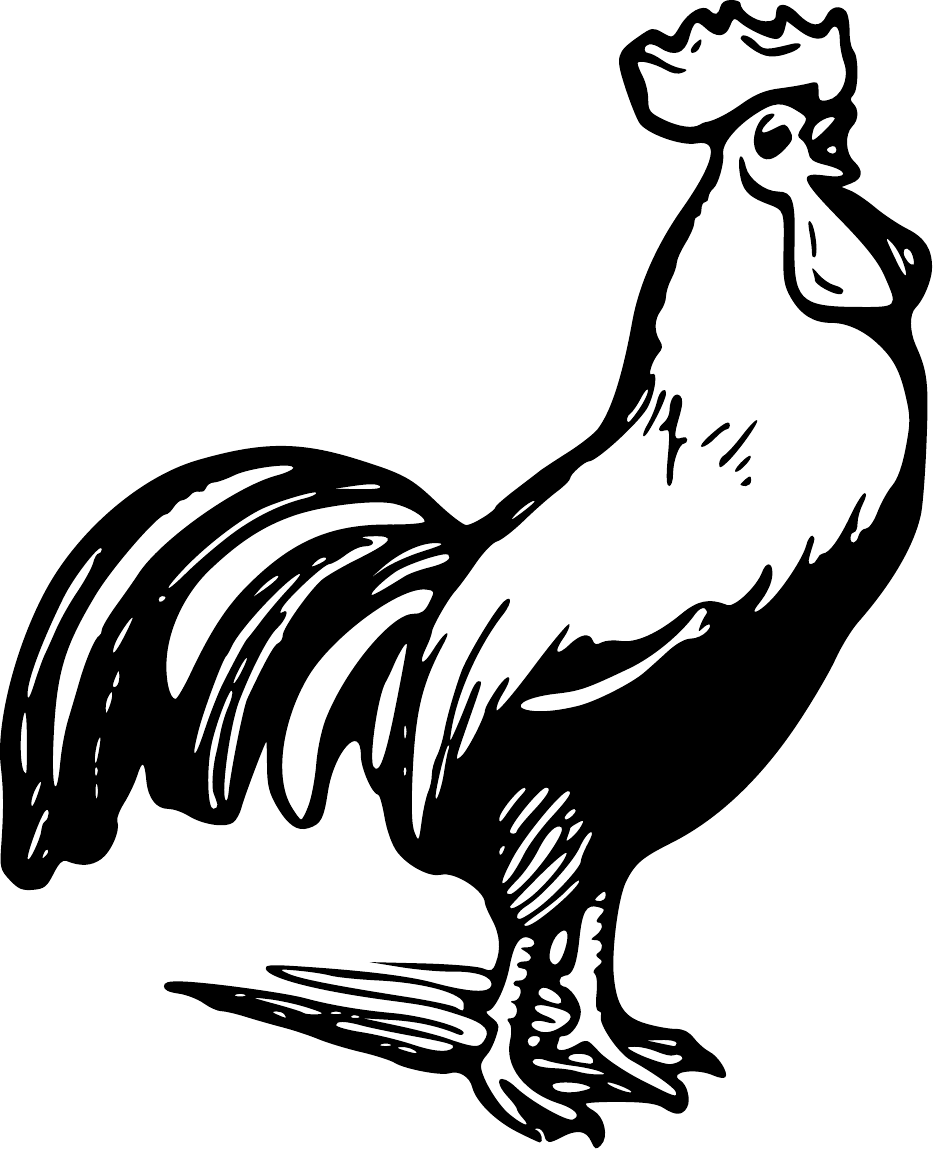}}
\newcommand{\CoqSymbol}{\raisebox{-.2ex}{\BareCoqSymbol\,}}
\newcommand{\Coqed}{\hfill\CoqSymbol}

\newdimen\zzfontsz

\newcounter{typerule}
\crefname{typerule}{rule}{rules}

\newcommand{\typeruleInt}[5]{%
	\def\thetyperule{#1}%
	\refstepcounter{typerule}%
	\label{tr:#4}%
  \ensuremath{\begin{array}{c}#5 \inference{#2}{#3}\end{array}}
}
\newcommand{\typerule}[4]{%
  \typeruleInt{#1}{#2}{#3}{#4}{\textsf{\scriptsize ({#1})} \\      }
}
\newcommand{\typerulenolabel}[3]{%
	\def\thetyperule{#1}%
	\refstepcounter{typerule}%
  \ensuremath{\begin{array}{c} \inference{#2}{#3}\end{array}}
}

\newcommand{\tmssafe}{\ensuremath\operatorname{tms}}
\newcommand{\smssafe}{\ensuremath\operatorname{sms}}
\newcommand{\mssafe}{\ensuremath\operatorname{ms}}
\newcommand{\ctsafe}{\ensuremath\operatorname{ct}}
\newcommand{\scctsafe}{\ensuremath\operatorname{scct}}
\newcommand{\msscctsafe}{\ensuremath\operatorname{mct}}
\newcommand{\sssafe}{\ensuremath\operatorname{ss}}
\newcommand{\specmssafe}{\ensuremath\operatorname{specms}}

\newcommand{\Ltms}{\ensuremath\src{L_{\tmssafe}}}
\newcommand{\Ltrg}{\ensuremath\trg{L}}
\newcommand{\Lms}{\ensuremath\irl{L_{\mssafe}}}
\newcommand{\Lscct}{\ensuremath\obj{L_{\scctsafe}}}
\MakeRobust{\mathghost}
\newcommand{\Lspec}{\ensuremath\ird{L_{\mathghost}}}

\newcommand{\event}[1][]{a#1}
\newcommand{\preevent}[1][]{b#1}

\newcommand{\emptyevent}{\ensuremath\varepsilon}
\newcommand{\trace}[1][]{\ensuremath\overline{a#1}}
\newcommand{\class}[1][]{\ensuremath\mb{C}}

\newcommand{\hole}[1]{\ensuremath{\left[#1\right]}}
\newcommand{\ev}[1]{\text{#1}}

\newcommand{\absterm}{\ensuremath\lightning{\kern-5.5pt}\lightning}

\newcommand{\subst}[2]{\ensuremath \hole{#1\text{ for }#2}}

\newcommand{\substlist}[1][]{\ensuremath \overline{\rho#1}}

\newcommand{\partialeval}[2]{\ensuremath \operatorname{\mathtt{mix}}(#1, #2)}

\newcommand{\natt}{\ensuremath\mb{N}\xspace}

\newcommand{\type}[1][]{\ensuremath\tau#1\xspace}

\newcommand{\wrapkeyword}[2][]{\ensuremath{#1{{#2}}}}
\newcommand{\expr}[1][]{e#1\xspace}

\newcommand{\valueexpr}[1][]{v#1\xspace}
\newcommand{\lbinop}[3][]{\ensuremath {#2}{#1{\oplus}}{#3}\xspace}
\newcommand{\lget}[3][]{\ensuremath #2{#1{[}}{#3}{#1{]}}\xspace}
\newcommand{\lset}[4][]{\ensuremath #2{#1{[}}{#3}{#1{]\leftarrow}}#4\xspace}
\newcommand{\lnew}[3][]{\ensuremath \wrapkeyword[#1]{new}\ #2\ {#1{[}}#3{#1{]}}\xspace}
\newcommand{\llet}[4][]{\ensuremath \wrapkeyword[#1]{let}\ #2 {#1{=}} #3\ \wrapkeyword[#1]{in}\ #4\xspace}
\newcommand{\ldelete}[2][]{\ensuremath \wrapkeyword[#1]{delete}\ #2\xspace}
\newcommand{\lreturn}[2][]{\ensuremath \wrapkeyword[#1]{return}\ #2\xspace}
\newcommand{\lcall}[3][]{\ensuremath \wrapkeyword[#1]{call}\ #2\ #3\xspace}
\newcommand{\lifz}[4][]{\ensuremath \wrapkeyword[#1]{ifz}\ #2\ \wrapkeyword[#1]{then}\ #3\ \wrapkeyword[#1]{else}\ #4\xspace}
\newcommand{\labort}[1][]{\ensuremath \wrapkeyword[#1]{abort()}\xspace}
\newcommand{\lispoisoned}[2][]{\ensuremath #2\ \wrapkeyword[#1]{is\ }{#1{\poisoned}}\xspace}
\newcommand{\lpair}[3][]{\ensuremath {#1{\langle}} #2 {#1{;}} #3 {#1{\rangle}} \xspace}

\newcommand{\lhast}[3][]{\ensuremath {#2}\ \wrapkeyword[#1]{has}\ #3 \xspace}
\newcommand{\lwrdoit}[2][]{\ensuremath \wrapkeyword[#1]{wrct}\ #2\xspace}
\newcommand{\lrddoit}[3][]{\ensuremath \wrapkeyword[#1]{rdct}\ #2\ \wrapkeyword[#1]{in}\ #3\xspace}

\newcommand{\lfunction}[4][]{\ensuremath\wrapkeyword[#1]{fn}\ {#2}\ {#3}\ {#1{:=}}\ #4\xspace}

\newcommand{\lbarrier}[1][]{\ensuremath \wrapkeyword[#1]{barrier}\xspace}

\newcommand{\rtp}[2]{\ensuremath\vdash{#1}:{#2}}
\newcommand{\rtpsim}[3]{\ensuremath\;\vdash^\forall_{{#3}}{#1}:{#2}}
\newcommand{\rtptau}[3]{\ensuremath\;\vdash^\exists_{{#3}}{#1}:{#2}}
\newcommand{\rtpsig}[3]{\ensuremath\;\vdash_{{#3}}{#1}:{#2}}
\newcommand{\ccbase}[1][]{\ensuremath\gamma{#1}}
\newcommand{\cc}[3][]{\ensuremath{\ccbase[#1]}^{#2}_{#3}\xspace}
\newcommand{\cca}{\ensuremath\cc{\Ltms}{\Ltrg}}
\newcommand{\ccb}{\ensuremath\cc{\Ltrg}{\Lms}}
\newcommand{\ccdce}{\ensuremath\cc[_{\gls{dce}}]{\Lms}{\Lms}}
\newcommand{\cccf}{\ensuremath\cc[_{\gls{cf}}]{\Lms}{\Lms}}
\newcommand{\ccscct}{\ensuremath\cc{\Lms}{\Lscct}}

\newcommand{\ccspec}{\ensuremath\cc{\Lscct}{\Lspec}}
\newcommand{\ccspecms}{\ccmssscct}
\newcommand{\ccmssscct}{\ensuremath\cc{\Ltms}{\Lspec}}

\newcommand{\wfctau}[2]{\ensuremath\vdash_{\mathit{wf}}{#1}:{#2}}
\newcommand{\wfcsig}[2]{\ensuremath\vdash_{\mathit{wf}}{#1}:{#2}}

\newcommand{\contextvar}[1][]{C#1}
\newcommand{\progvar}[1][]{p#1}
\newcommand{\wholeprogvar}[1][]{w#1}
\renewcommand{\class}[1][]{\mathbb{C}#1}
\newcommand{\link}[2]{\ensuremath\operatorname{link}\left({#1};{#2}\right)}

\newcommand{\sat}[2]{\ensuremath\vdash{#1}:{#2}}
\newcommand{\rsat}[2]{\ensuremath\vdash_R{#1}:{#2}}

\newcommand{\securitytag}[1][]{\ensuremath s#1}
\newcommand{\sandboxtag}[1][]{t#1}
\newcommand{\ctx}{\text{ctx}}
\newcommand{\comp}{\text{comp}}
\newcommand{\loc}[1][]{\ensuremath l#1}
\newcommand{\poison}{\ensuremath\rho}
\newcommand{\poisoned}{\ensuremath\text{\Biohazard}}

\newcommand{\library}[1][]{\ensuremath\Xi#1}

\newcommand{\cfstate}[1][]{\ensuremath\Psi#1}
\newcommand{\memstate}[1][]{\ensuremath\Phi#1}

\newcommand{\rtt}[2]{\ensuremath #1 \triangleright #2}

\newcommand{\runtimetermvar}[1][]{r#1}

\newcommand{\stepton}[4][n]{\ensuremath{#2}\xrightarrow{#4}{}{\kern-3.5pt}^{#1}\xspace{#3}\xspace}

\newcommand{\ghoststepto}[3]{\ensuremath{#1}\xrightswishingghost{#3}\xspace{#2}\xspace}
\newcommand{\pstepto}[3]{\ensuremath{#1}\xrightarrow{#3}_p{}\xspace{#2}\xspace}

\newcommand{\estepton}[4][n]{\ensuremath{#2}\xrightarrow{#4}_{\operatorname{ectx}}{}{\kern-14.5pt}^{#1\ \ \ \;}\xspace{#3}\xspace}
\newcommand{\progstepto}[3]{\ensuremath{#1}\xRightarrow{#3}{#2}}

\allowdisplaybreaks

\Crefname{exampleenv}{Example}{Examples}

\theoremstyle{definition}
\newtheorem{exampleenv}{Example}[section]
\newtheorem{lemma}{Lemma}[section]
\newtheorem{theorem}{Theorem}[section]
\newtheorem{corollary}{Corollary}[section]
\newtheorem{definition}{Definition}[section]

\newacronym{rtp}{RTP}{Robust Trace-Property Preservation}
\newacronym{tms}{TMS}{Temporal Memory Safety}
\newacronym{sms}{SMS}{Spatial Memory Safety}
\newacronym{ms}{MS}{Memory Safety}
\newacronym{specms}{SpecMS}{Speculative Memory Safety}
\newacronym{cct}{CCT}{Cryptographic Constant Time}
\newacronym{scct}{sCCT}{Strict Cryptographic Constant Time}
\newacronym{ss}{SS}{Speculative Safety}
\newacronym{dce}{DCE}{Dead Code Elimination}
\newacronym{cf}{CF}{Constant Folding}

\newacronym{msscct}{MCT}{Memory Safety and Strict Cryptographic Constant Time}
\newacronym{mssscct}{MCT$\mathghost$}{Memory Safety and Strict Cryptographic Constant Time with Speculation}

\makeglossaries

\AtBeginDocument{%
  \providecommand\BibTeX{{%
    \normalfont B\kern-0.5em{\scshape i\kern-0.25em b}\kern-0.8em\TeX}}}
\makeatletter
\newcommand{\linebreakand}{%
  \end{@IEEEauthorhalign}
  \hfill\mbox{}\par
  \mbox{}\hfill\begin{@IEEEauthorhalign}
}
\makeatother

\begin{document}

\title{
  Secure Composition 
  \\ 
  of Robust and Optimising Compilers
}

\author{
Matthis Kruse
\and
Michael Backes
\and
Marco Patrignani
}

\maketitle

\begin{abstract}
To ensure that secure applications do not leak their secrets, they are required to uphold several security properties such as spatial and temporal memory safety, cryptographic constant time, as well as speculative safety.
Existing work shows how to enforce these properties individually, in an architecture-independent way, by using secure compiler passes that each focus on an individual property.
Unfortunately, given two secure compiler passes that each preserve a possibly different security property, it is unclear what kind of security property is preserved by the composition of those secure compiler passes.
This paper is the first to study what security properties are preserved across the composition of different secure compiler passes.
Starting from a general theory of property composition for security-relevant properties (such as the aforementioned ones), this paper formalises a theory of composition of secure compilers.
Then, it showcases this theory on a secure multi-pass compiler that preserves the aforementioned security-relevant properties.
Crucially, this paper derives the security of the multi-pass compiler from the composition of the security properties preserved by its individual passes, which include security-preserving as well as optimisation passes.
From an engineering perspective, this is the desirable approach to building secure compilers.
\end{abstract}

\section{Introduction\pages{4}}\label{sec:introduction}

\gls*{ms} is a security property obtained by composing \gls*{sms}, which ensures array accesses are all within bounds, and \gls*{tms}, which ensures pointers are only used when they are valid~\cite{azevedo2018meaningsofms,jim2002cyclone,necula2005ccured,nagarakatte2010cets,nagarakatte2009soft,akritidis2009baggy,michael2023mswasm}.
\gls*{cct} is a security property that ensures sensitive data is not leaked via timing side-channels~\cite{kocher1996timing} and \gls*{ss} is a security property that enforces the same but under a speculative semantics~\cite{guarnieri2018spectector,fabian2022automatic} that captures speculative execution attacks such as spectre~\cite{kocher2019spectre}.
Together, \gls*{sms}, \gls*{tms}, \gls*{cct} and \gls*{ss}, yield \gls*{specms}, which is the gold standard of security properties for secure applications.
Programs attaining \gls*{specms} do not leak sensitive data either through erroneous memory accesses, nor through timing side-channels, even under speculative execution.
\Cref{ex:strncpy} below discusses how these security properties can be enforced by compiler passes~\cite{bond2017vale,almeida2017jasmin}, to ensure programmers need not be aware of the architectural details of where their code will run.

\begin{exampleenv}[strncpy]\label{ex:strncpy}
Consider the following \texttt{C} function \texttt{strncpy} that copies a null-terminated string \texttt{src} into \texttt{dst} up to a length of \texttt{n} characters.
This function is subject to a subtle \gls*{sms} vulnerability: the bounds check \texttt{i < n} should happen {\it before} the access to memory location \texttt{src[i]}: otherwise
the memory location past the last element is leaked to an attacker.
\begin{lstlisting}[language=c,basicstyle=\small\ttfamily,morekeywords={size_t}]
void strncpy(size_t n, char *dst, char *src) {
  for(size_t i = 0; src[i] != '\0' && i < n; ++i) {
    dst[i] = src[i];
  }
}
\end{lstlisting}
To prevent this vulnerability, one can use a compilation pass that enforces \gls*{sms}, such as Softbounds~\cite{nagarakatte2009soft} or BaggyBounds~\cite{akritidis2009baggy}.
The most na\"ive solution in this case is to insert bounds-checks in front of every access to memory.

Because of timing attacks, addressing \gls*{sms} is not enough to make \texttt{strncpy} secure.
In fact, the loop can terminate early, as soon as the string-terminating character \texttt{'\textbackslash 0'} is encountered, thus making program execution time proportional to the length of the array pointed by \texttt{src}, and violating the \gls*{cct} property.
Also in this case there exist compiler passes that can rewrite such programs into \gls*{cct} ones~\cite{cauligi2019fact}.
Finally, even with these precautions, code is not run in isolation, so a malicious attacker could supply code that interacts with \texttt{strncpy} and trigger a violation of either \gls*{ms} or \gls*{cct} by calling \texttt{strncpy} with an argument for \texttt{src} that points to uninitialised memory.
This would, in turn, trigger a series of reads from uninitialised memory, which is an immediate \gls*{ms} violation with devastating real-world consequences~\cite{uninit-0,uninit-1,uninit-2,uninit-3,uninit-4}.
\end{exampleenv}

Whether or not compiler passes enforce certain security properties, it is important for a secure compiler to consider partial programs that interact with potentially malicious code, since the latter may lead to, e.g., memory-safety issues in the considered partial program.
Robust compilers~\cite{abate2019jour} are a form of secure compilers that preserve security properties even in the presence of arbitrary attackers interacting with compiled code.
Thus, robust compilers can be used to prevent vulnerabilities resulting from uninitialised memory (as well as many other ones), e.g., by targeting capability-based languages such as CHERI~\cite{woodruff2014CHERI}, Arm Morello~\cite{arm-morello}, or MSWasm~\cite{michael2023mswasm}, where the compiler relies on capabilities to check that pointers are always initialised.

Unfortunately, given secure compiler passes that each preserve a possibly different security property, there is no way to tell what kind of security property will the composition of those secure compilers preserve.
Worse, without a framework for composing secure compiler passes, it is not possible to enable separation of concerns, e.g., to have a secure compilation pass that ensures \gls*{ms} that is developed independently of another secure pass for \gls*{cct}, that is developed independently of other passes, such as optimisation ones.

This paper introduces a framework for reasoning about the composition of secure and optimising compiler passes %
and it showcases the power of this framework by instantiating it on a multi-pass compilation chain.
To this end, this paper first discusses how to compose security properties, such as \gls*{tms} and \gls*{sms} into \gls*{ms}, and then adds \gls*{cct} as well as \gls*{ss} into the mix to obtain \gls*{specms}.
The paper then defines several secure compiler passes, where each is either preserving a different security property from the list above
or performing a security-preserving optimisation.
Finally, this paper shows that composing these secure compiler passes into a multi-pass compilation chain results in the end-to-end (robust) preservation of \gls*{specms}.
Crucially, this paper derives the security of the multi-pass compiler from the composition of the security properties preserved by its individual passes.
This result showcases how the framework allows the kind of formal security reasoning that compiler writers already want (and already do), obtaining precise, compositional security reasoning while providing minimal (and modular) proof effort.
In summary, this paper makes the following contributions:
\begin{asparaitem}[$\blacktriangleright$]
  \item %
        This paper takes the secure compilation framework of Abate et al.~\cite{abate2021extacc} and extends it for compositionality. (\Cref{sec:sequential}).
        This paper proves that starting from two compilers that preserve two (possibly distinct) properties, their composition preserves roughly the intersection of those properties.
        Then, this paper identifies which conditions make the composition of secure compilers meaningful from a security perspective, and which other conditions allow passes to be swapped without losing security meaning.

  \item %
        This paper presents a case-study with five programming languages showcasing the previous contribution (\Cref{sec:casestud:defs,sec:casestud:rtp}).
        To this end, it presents a compilation chain consisting of six passes that ultimately preserve \gls*{specms} by means of composing four secure passes that \emph{individually} preserve \gls*{tms}, \gls*{sms}, \gls*{scct}---an enforceable version of \gls*{cct}---, and \gls*{ss}, and two passes which are well-known optimisations \gls*{dce} and \gls*{cf}.
        The formalisation of this case study showcases the power of the presented framework: the divide-and-conquer approach to software engineering is a viable strategy even for the development of secure compilers.

  \item %
  		This paper provides formal proofs of the security properties preserved by each of the presented passes. 
  		Additionally this paper analyses each pass in terms of their compatibility with each other (\Cref{sec:formalities}), proving how does each pass fulfill the conditions that make their composition meaningful from a security standpoint.

  \item The key contributions of this paper are formalised in the Coq proof assistant and the paper indicates this with \CoqSymbol.
\end{asparaitem}
This paper starts by introducing relevant notions of security properties and secure compilation (\Cref{sec:background}),
and discusses related work (\Cref{sec:relwork}) before concluding (\Cref{sec:concl}).

The omitted formal details, lemmas and proofs, as well as the Coq formalisation are available as supplementary material.

\section{Background: Properties and Secure Compilers\pages{1}}\label{sec:background}

To introduce the security argument of this paper, this section defines (security) properties, their satisfaction, and their robust satisfaction (i.e., satisfaction in the presence of an active attacker; \Cref{subsec:bg:tprop}).
Then, borrowing from existing work~\cite{abate2019jour,abate2021extacc}, this section introduces secure compilers as compilers that preserve robust property satisfaction (\Cref{subsec:bg:rtp}).

\subsection{Properties and (Robust) Satisfaction}\label{subsec:bg:tprop}

This paper employs the security model where programs are written in a language whose semantics emits events $\event$.
Events include security-relevant actions (e.g., reading from and writing to memory, as detailed in \Cref{sec:compprop}) and the unobservable event $\emptyevent$.
As programs execute, their emitted events are concatenated in traces $\trace$, which serve as the description of the behaviour of a program.%
\footnote{
Throughout the paper, sequences are indicated with an overbar (i.e., $\trace$), empty sequences with $\hole{\cdot}$, and concatenation of sequences $\trace[_{1}],\trace[_{2}]$ as $\trace[_{1}]\cdot\trace[_{2}]$.
Prepending element $\event$ to a sequence $\trace$ uses the same notation: $\event\cdot\trace$.
}

Properties $\pi$ are sets of traces of admissible program behaviours, ascribing what said property considers valid.
The set of all properties can be partitioned into different {\em classes} ($\class$), i.e., safety, liveness, and neither safety nor liveness~\cite{clarkson2008hyper}, so a class is simply a set of properties.
The compositionality framework (\Cref{sec:sequential}) presented in this paper considers arbitrary classes, while the case-study (\Cref{sec:casestud:rtp}) fixes them to concrete instances of safety properties, since it is decidable whether a trace satisfies a safety property with just a finite trace (i.e., a \emph{prefix}).
As an example, consider the trace:
$$\ev{Dealloc\ \loc}\cdot\ev{Read\ \loc\ 1729}\cdot\dots$$
which describes the interaction with a memory where the deallocation of an address $\loc$ precedes a read (of some value $1729$) at that address in memory.
This program behaviour is insecure w.r.t. a canonical notion of (temporal) memory safety dictating no use-after-frees of pointers~\cite{nagarakatte2010cets,azevedo2018meaningsofms}, because it reads from a memory location that was freed already.
The previous finite trace prefix is enough to decide that the trace does not satisfy \gls*{tms} and there is no way to append events to this prefix which would result in the trace being admissible.

In the following, the execution of a whole program $\wholeprogvar$ that terminates in state $\runtimetermvar$ according to the language semantics and produces trace $\trace$ is written as $\progstepto{\wholeprogvar}{\runtimetermvar}{\trace}$.
With this, we defined property satisfaction as follows:
\bul{a whole program $\wholeprogvar$ satisfies a property $\pi$} iff \iul{$\wholeprogvar$ yields a trace $\trace$} such that \oul{$\trace$ satisfies $\pi$}.

\begin{definition}[Property Satisfaction]\label{def:propsat}
$\;$ 

  \begin{nscenter}
    \bul{$\sat{\wholeprogvar}{\pi}$}
    $\isdef$
    if \iul{$\forall\runtimetermvar\ \trace,\progstepto{\wholeprogvar}{\runtimetermvar}{\trace}$},
    then \oul{$\trace\in\pi$}.
  \end{nscenter}
\end{definition}

Property satisfaction is defined on whole programs, i.e., programs without missing definitions.
Thus, from a security perspective, \Cref{def:propsat} considers only a passive attacker model, where the attacker observes the execution and, e.g., retrieves secrets from that.
To consider a stronger model similarly to what existing work does~\cite{abate2019jour,abate2021extacc,maffeis2008code-carrying,gordon2003authenticity,fournet2007authorization,bengtson2011refine,backes2014uniontyps,michael2023mswasm,swasey2017robust,sammler2019benefits}, we extend the concept of satisfaction with {\em robustness}.
Robust satisfaction considers partial programs $\progvar$, i.e., components with missing imports, which cannot run until said imports are fulfilled.
To remedy this, {\em linking} takes two partial programs $\progvar[_{1}],\progvar[_{2}]$ and produces a whole program $\wholeprogvar$, i.e., $\link{\progvar[_{1}]}{\progvar[_{2}]}=\wholeprogvar$.
As typically done in works that consider the execution of partial programs~\cite{abate2019jour,devriese2018parametricity,patrignani2021rsc,korashy2021capableptrs,strydonck2019lincap,devriese2017modular,bowman2015noninterference,ahmed2011equivcps,patterson2017linkingtyps},
this paper assumes that whole programs are the result of linking partial programs referred to as {\em context} ($\ctx$) and {\em component} ($\comp$).
The context is an arbitrary program and thus has the role of an {\em attacker} that can interact with the component by means of any features the programming language has, and the component is what is security-relevant.
With this, \Thmref{def:propsat} can be extended as follows: for \bul{a component $\progvar$ to robustly satisfy a property $\pi$}, take an \iul{attacker context $\contextvar$ and link it with $\progvar$}, \oul{the resulting whole program must satisfy $\pi$}.

\begin{definition}[Robust Satisfaction]\label{def:proprsat}
  $\;$ 

  \begin{nscenter}
  \bul{$\rsat{\progvar}{\pi}$}
  $\isdef$ \iul{$\forall \contextvar\ \wholeprogvar$, if $\link{\contextvar}{\progvar}=\wholeprogvar$}, then \oul{$\sat{\wholeprogvar}{\pi}$}.
  \end{nscenter}
\end{definition}

\subsection{Secure Compilers}\label{subsec:bg:rtp}

A {\em compiler} ($\cc{\src{L}}{\trg{L}}$) translates syntactic descriptions of programs from a {\em source} ($\src{L}$) into a {\em target} ($\trg{L}$) programming language.
This translation is considered {\em correct} if it is semantics-preserving.
That is, for a whole program $\src{\wholeprogvar}$, the compiler should relate the $\src{L}$ semantics of $\src{\wholeprogvar}$ with the semantics of $\trg{L}$ of the compiled counterpart of $\src{p}$ in such a way that they are ``compatible''.
Unfortunately, correct compilers may be insecure compilers~\cite{patrignani2019survey,kennedy2006secure.net,abadi1999protect,ahmed2018dagstuhl} and programs translated by insecure compilers can violate security properties that the programmer assumes to hold.
This is why {\em robust preservation} is a good candidate as a compiler-level security property~\cite{abate2019jour}.

This paper uses a general notion of robust preservation~\cite{abate2021extacc} that considers compilers that use languages with different trace models. 
To this end, considering a source trace $\src{\trace}$ and a target trace $\trg{\trace}$, there is a relation ($\sim$) describing the effect of a corresponding compiler (see \Cref{sec:casestud:rtp}). 
This relation induces the following two projection functions~\cite{abate2021extacc}: (1) the \emph{existential image} $\tau_\sim\left(\src{\pi}\right)$ and (2) the \emph{universal image} $\sigma_\sim\left(\trg{\pi}\right)$.
These projections map source-level (resp. target-level) properties to target-level (resp. source-level) properties in a way that identifies the ``same'' property across languages. 
The case study of this paper uses the universal image, since some considered properties, such as \gls*{ss}, are not definable in a higher-level language that, e.g., does not model speculation.
\begin{definition}[Universal Image]
\label{def:universal:img}\label{def:sigma}
  \[ 
    \sigma_\sim\left(\trg{\pi}\right) := 
      \left\{ 
        \src{\trace} \mid \forall \trg{\trace}\ldotp \text{if }\src{\trace}\sim\trg{\trace}, \text{ then } \trg{\trace}\in\trg{\pi} 
      \right\}
  \]
\end{definition}
\noindent
With this projection function, we define a more general version of robust preservation as follows~\cite{abate2021extacc}.
A \bul{compiler $\cc{\src{L}}{\trg{L}}$ robustly preserves a class of target properties $\trg{\class}$}, if for any \rul{property $\trg{\pi}$ of class $\trg{\class}$ and programs $\src{p}$}, where \iul{$\src{\progvar}$ robustly satisfies $\sigma_\sim\left(\trg{\pi}\right)$}, \oul{the compilation of $\src{\progvar}$, we have that $\cc{\src{L}}{\trg{L}}\left(\src{p}\right)$ robustly satisfies $\trg{\pi}$}.

\begin{definition}[Robust Preservation with $\sigma_\sim$]\label{def:rtp:sigma}
  \bul{$\rtpsig{\cc{\src{L}}{\trg{L}}}{\trg{\class}}{\sim}$}
  $\isdef$
  \rul{$\forall \trg{\pi}\in\trg{\class}, \src{p}\in\src{L},$} if \iul{$\rsat{\src{\progvar}}{\sigma_\sim\left(\trg{\pi}\right)}$}, then \oul{$\rsat{\cc{\src{L}}{\trg{L}}\left(\src{p}\right)}{\trg{\pi}}$}.
\end{definition}
Note that a class of properties $\class$ can represent just one property $\pi$ by lifting~\cite{clarkson2008hyper} that property to sets of properties, i.e., use the powerset of $\pi$ instead of $\pi$ itself.
Because of this, this paper may write $\rtptau{\cc{\src{L}}{\trg{L}}}{\trg{\pi}}{\sim}$, even though $\trg{\pi}$ is a property and not a class.
A similar construction can be used to the projection functions (see \Cref{def:universal:img}) by applying them to the lifted version of $\trg{\pi}$ instead of just $\trg{\pi}$.

In case the trace model is the same for both source and target programs (and thus $\sim$ is equality), we obtain~\cite{abate2019jour}:
\begin{definition}[Robust Preservation]\label{def:rtp}
  $\;$

  {$\rtp{\cc{\src{L}}{\trg{L}}}{\class}$}
  $\isdef$
  {$\forall \pi\in\class, \src{p}\in\src{L},$} if {$\rsat{\src{\progvar}}{\pi}$}, then {$\rsat{\cc{\src{L}}{\trg{L}}\left(\src{p}\right)}{\pi}$}.
\end{definition}

Examples of compilers fulfilling \Cref{def:rtp} exist in the literature~\cite{korashy2022secureptrs,korashy2021capableptrs,abate2021extacc,abate2019jour,patrignani2021rsc}.
For example, SecurePtrs~\cite{korashy2022secureptrs} gives a compiler that robustly preserves all safety properties for a C-like language to an assembly-like language. 
As another example, even though it is not strictly satisfying \Cref{def:rtp}, the FaCT~\cite{cauligi2019fact} compiler preserves the \gls*{cct} property for a C-like language with constant-time primitives, e.g., \texttt{ctselect} for branching. %
Throughout this work, it is assumed that FaCT satisfies \Cref{def:rtp}.

\section{Secure Composition}\label{sec:sequential}

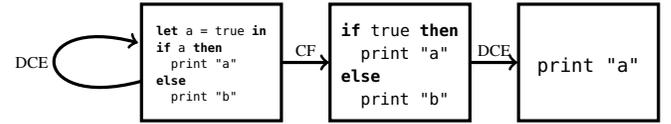
\begin{figure}[!ht]
  \begin{center}
    \begin{tikzpicture}[scale=0.62, every node/.style={transform shape}]
      \node[minimum width=3cm,minimum height=2.5cm,draw=black,very thick,align=left] (progunopt) {\begin{lstlisting}[language=ml,basicstyle=\small\ttfamily]
let a = true in
if a then
  print "a"
else
  print "b"
\end{lstlisting}};
      \node[minimum width=3cm,minimum height=2.5cm,draw=black,very thick,align=left,right=1.0 of progunopt] (progoptcf) {\begin{lstlisting}[language=ml,basicstyle=\large\ttfamily]
if true then
  print "a"
else
  print "b"
\end{lstlisting}};
      \node[minimum width=3cm,minimum height=2.5cm,draw=black,very thick,align=left,right=1.0 of progoptcf] (progoptcfdce) {\begin{lstlisting}[language=ml,basicstyle=\Large\ttfamily]
print "a"
\end{lstlisting}};%
      \draw[->,very thick] (progunopt) edge[loop left,left] node {\gls*{dce}} (progunopt);
      \draw[->,very thick] (progunopt) edge[sloped,above] node {\gls*{cf}} (progoptcf);
      \draw[->,very thick] (progoptcf) edge[sloped,above] node {\gls*{dce}} (progoptcfdce);
    \end{tikzpicture}
  \end{center}
  \caption{Example program where the level of optimisations differ for one pass of applying \gls*{cf} and \gls*{dce} in any order. %
  Every edge is a compilation pass and the label on the edge states what the pass does, i.e., \gls*{cf} or \gls*{dce}. %
  The source code in the nodes is a glorified compiler intermediate representation and the code gets more optimised towards the right hand side of the figure.}\label{fig:cfdceex}
\end{figure}

Notably, real-world compilation chains also perform a series of (sequential) passes whose main purpose is not necessarily to translate from one language to another, but to, e.g., optimise the code or enforce a certain property.
Both examples can be seen in practice, e.g.,~\cite{nagarakatte2009soft,nagarakatte2010cets,akritidis2009baggy,wegman1991ccp,manjikian1997fusion} and many more.
Consider the following two LLVM optimisation passes: \gls*{cf}, which rewrites constant expressions to the constant they evaluate to, and \gls*{dce}, which removes dead code by rewriting conditional branches.
The order in which \gls*{cf} and \gls*{dce} are performed influences the final result of the compilation (see \Cref{fig:cfdceex}).
This {\em phase ordering problem} is well--known in literature and a practical solution is to simply perform a fixpoint iteration of the optimisation pipeline~\cite{click1995combining}.

\smallskip

To analyse the security of compilation passes and their interaction within a compilation pipeline, we rely on a few key notions: the definition of a trace relation, and the definition of when is a trace relation well-formed with respect to a class.
Consider traces $\src{\trace}$ and $\trg{\trace}$ as well as two trace relations $\sim_1$ and $\sim_2$. 
The traces are related $\src{\trace}\sim_1\bullet\sim_2\trg{\trace}$ if there is another trace $\irl{\trace}$ such that $\src{\trace}\sim_1\irl{\trace}$ and $\irl{\trace}\sim_2\trg{\trace}$.
\bul{A relation $\sim$ is well-formed w.r.t. a class of target-level properties $\trg{\class}$} iff \rul{the universal image preserves set membership}.
\begin{definition}[Well-formedness of $\sim$ for a Class $\trg{\class}$]\label{def:wfc:sig:tracerel}

  \begin{nscenter}
  \noindent
  \text{\bul{$\wfcsig{\sim}{\trg{\class}}$}} \isdef \text{\rul{$\forall \trg{\pi}\in\trg{\class}, \sigma_\sim(\trg{\pi})\in\sigma_\sim(\trg{\class})$}}
  \end{nscenter}
\end{definition}

We can now state our main result: secure compilers in the robust compilation framework~\cite{abate2021extacc} compose {\em sequentially}. 

Let \bul{$\cc{\src{L}}{\trg{L}}$ robustly preserve the class $\sigma_{\sim_2}\left(\trg{\class[_1]}\right)$ under $\sim_1$} and let \rul{$\cc{\trg{L}}{\obj{L}}$ robustly preserve the class $\trg{\class[_2]}$ under $\sim_2$}.
Then, \iul{when the cross-language relation $\sim_2$ is well-formed w.r.t. class $\trg{\class[_1]}$}, it follows that \oul{the composed compiler $\cc{\src{L}}{\trg{L}}\circ\cc{\trg{L}}{\obj{L}}$ robustly preserves the intersection of classes $\trg{\class[_1]}\cap\trg{\class[_2]}$ under $\sim_1\bullet\sim_2$}.
\begin{theorem}[Composition of Secure Compilers w.r.t. $\sigma$]\label{thm:rtpsim:sig}
  $\;$ 

  If \bul{$\rtpsig{\cc{\src{L}}{\trg{L}}}{\tilde{\sigma}_{\sim_2}\left(\trg{\class[_{1}]}\right)}{\sim_1}$}, \rul{$\rtpsig{\cc{\trg{L}}{\obj{L}}}{\trg{\class[_2]}}{\sim_2}$}, and \iul{$\wfcsig{\sim_2}{\trg{\class[_1]}}$}, \\ then \oul{$\rtpsig{\cc{\src{L}}{\trg{L}}\circ\cc{\trg{L}}{\obj{L}}}{\trg{\class[_{1}]}\cap\trg{\class[_{2}]}}{\sim_1\bullet\sim_2}$}. \Coqed
\end{theorem}

Since the composition of secure compilers is again a secure compiler, the theorems generalise to a whole chain of $n$ secure compilers. 
\Cref{thm:rtpsim:sig} can also be stated for the existential image $\tau_\sim\left(\src{\pi}\right)$, but in the interest of space that result has been moved to the appendix.
Crucially, \Cref{thm:rtpsim:sig} also holds for \emph{classes of hyperproperties}, and thus, compilers that robustly preserve hyperproperties can be composed with each other as well as with compilers that robustly preserve properties.

If we take SecurePtrs and FaCT from \Cref{subsec:bg:rtp} and compose them according to \Cref{thm:rtpsim:sig}, we obtain a compiler that robustly preserves the intersection of safety properties and the \gls*{cct} hyperproperty. 
That is, for a source component that robustly satisfies any set of safety properties {\em and} \gls*{cct}, the compiled target component also robustly satisfies the same set of safety properties and \gls*{cct}.

Compiler engineers typically try to find an order of optimisations that yields well-optimised programs for either code size~\cite{cooper1999geneticphases} or performance~\cite{kulkarni2006exhaustivephase}.
\Cref{corr:swappable} justifies that any such order of compilation passes is valid w.r.t. security, as long as the trace-relations have no effect on the respective classes.

So, given two compilation passes \bul{$\cc[_{1}]{\trg{L}}{\trg{L}}$, $\cc[_{2}]{\trg{L}}{\trg{L}}$, both robustly preserving class $\trg{\class[_{1}]}$ or $\trg{\class[_{2}]}$, respectively}, \rul{their corresponding well-formed trace-relations}, \iul{and indifference of the classes with respect to these trace relations}, \oul{for any order of their composition, the composed compiler robustly preserves the intersection of $\trg{\class[_{1}]}$ and $\trg{\class[_{2}]}$}.

\begin{corollary}[Swapping Secure Compiler Passes]\label{corr:swappable}
  $\;$ 

  If \bul{$\rtpsig{\cc[_{1}]{\trg{L}}{\trg{L}}}{\trg{\class[_{1}]}}{\sim_1}$ and $\rtpsig{\cc[_{2}]{\trg{L}}{\trg{L}}}{\trg{\class[_{2}]}}{\sim_2}$}, %
  \rul{$\wfcsig{\sim_1}{\trg{\class[_2]}}$ and $\wfcsig{\sim_2}{\trg{\class[_1]}}$}, %
  and \iul{$\tilde{\sigma}_{\sim_2}\left(\trg{\class[_1]}\right)=\trg{\class[_1]}$ as well as $\tilde{\sigma}_{\sim_1}\left(\trg{\class[_2]}\right)=\trg{\class[_2]}$},
  then \oul{$\rtpsig{\cc[_{1}]{\trg{L}}{\trg{L}}\circ\cc[_{2}]{\trg{L}}{\trg{L}}}{\trg{\class[_{1}]}\cap\trg{\class[_{2}]}}{\sim_1\circ\sim_2}$ and $\rtpsig{\cc[_{2}]{\trg{L}}{\trg{L}}\circ\cc[_{1}]{\trg{L}}{\trg{L}}}{\trg{\class[_{2}]}\cap\trg{\class[_{1}]}}{\sim_2\circ\sim_1}$}. \Coqed
\end{corollary}

Coming back to the example composing SecurePtrs with FaCT, it is likely the case that \Cref{corr:swappable} is not applicable.
While first running SecurePtrs and then FaCT should be fine, the other direction has potential security hazards, since the SecurePtrs compiler does not account for cryptographic-constant time primitives, such as \texttt{ctselect}.

\subsection{Secure Compiler Composition with Same Trace Models}
When the cross-language trace relation is an equality, \Cref{thm:rtpsim:sig} collapses:
Given \bul{$\cc{\src{L}}{\trg{L}}$ robustly preserves $\class[_{1}]$} and \rul{$\cc{\trg{L}}{\obj{L}}$ robustly preserves $\class[_{2}]$}, it follows that \oul{their sequential composition $\cc{\src{L}}{\trg{L}}\circ\cc{\trg{L}}{\obj{L}}$ robustly preserves the intersection of classes $\class[_{1}]$ and $\class[_{2}]$}.

\begin{corollary}[Composition of Secure Compilers]\label{corr:rtp}
  $\;$ 

  If \bul{$\rtp{\cc{\src{L}}{\trg{L}}}{\class[_{1}]}$} and \rul{$\rtp{\cc{\trg{L}}{\obj{L}}}{\class[_{2}]}$}, then \oul{$\rtp{\cc{\src{L}}{\trg{L}}\circ\cc{\trg{L}}{\obj{L}}}{\class[_{1}]\cap\class[_{2}]}$}. \Coqed
\end{corollary}

\Cref{corr:rtp} provides an easy way to compose secure compilers without well-formedness of trace relations. 
However, while \Cref{thm:rtpsim:sig} explicitly requires that $\sim_1$ is well-formed w.r.t. $\trg{\class[_2]}$, if $\sim_2$ is not well-formed w.r.t. $\trg{\class[_1]}$, care must be taken. 
This is further discussed in \Cref{subsec:compatsecpasses}.

We can also obtain a specialised version of \Cref{corr:swappable}:

\begin{corollary}[Swapping Secure Compiler Passes]\label{corr:swappable:one}
  $\;$ 

  If {$\rtp{\cc[_{1}]{\trg{L}}{\trg{L}}}{\class[_{1}]}$ and $\rtp{\cc[_{2}]{\trg{L}}{\trg{L}}}{\class[_{2}]}$}, then {$\rtp{\cc[_{1}]{\trg{L}}{\trg{L}}\circ\cc[_{2}]{\trg{L}}{\trg{L}}}{\class[_{1}]\cap\class[_{2}]}$ and $\rtp{\cc[_{2}]{\trg{L}}{\trg{L}}\circ\cc[_{1}]{\trg{L}}{\trg{L}}}{\class[_{2}]\cap\class[_{1}]}$}. \Coqed
\end{corollary}

\section{Security Properties Formalisation \& Composition\pages{1.5}}\label{sec:compprop}

This section introduces trace properties of interest for this paper: \gls*{tms}, \gls*{sms}, \gls*{ms}, \gls*{scct}, and \gls*{ss}.
These properties are of practical importance (as mentioned in \Cref{sec:introduction}) and also of interest in the case study presented later (\Cref{sec:casestud:defs,sec:casestud:rtp}). 
This section presents all of them, despite the fact that they are inspired by existing work, in order to showcase all that is required for a formal proof of security for a realistic compilation toolchain.
The technical report defines monitors for each of the presented properties.
Monitors refine each property and have a key tole in the proofs of this paper.

\subsection{A Trace Model for Memory Safety}\label{subsec:basic:memsafety:tracemodel}

For simple memory safety composed of temporal and spatial memory safety, the trace model defines events ($\event[_{\mssafe}]$) as either the empty event ($\emptyevent$), a crash ($\lightning$), or a base-event ($\preevent[_{\mssafe}]$).

\vspace{-1.0em}
\[
  \begin{array}{rcll}
    \text{(Base-Events)}&\preevent[_{\mssafe}] &:=& \ev{Alloc \loc\ n} \mid \ev{Dealloc \loc} \mid \ev{Use \loc\ n} \\
    \text{(Events)}&\event[_{\mssafe}] &:=& \preevent[_{\mssafe}] \mid \emptyevent \mid \lightning \\ 
  \end{array}
\]

Base-events describe the actual kind of event that happened.
For the basic memory-safety properties, these are three variants:
First, the allocation event ($\ev{Alloc\ \loc\ n}$) that fires whenever a program claims $n$ cells of memory and stores them at address $\loc$, where addresses are assumed to be unique.
Second, deallocation ($\ev{Dealloc\ \loc}$) announces that the object at location $\loc$ is freed.
Third, an event to describe reads from and writes to the $n$-th memory cell from address $\loc$ ($\ev{Use\ \loc\ n}$).

\subsubsection{Temporal Memory Safety}

\gls*{tms}~\cite{nagarakatte2010cets} is a safety property that describes that an unallocated object must not be (re-)used.

\begin{definition}[\glsfirst*{tms}]\label{def:trace:tmsdef}
  $$
  \tmssafe:=\left\{\trace[_{\mssafe}] \left| \begin{array}{rcl}
    \ev{Alloc\ \loc\ n}&\le_{\trace[_{\mssafe}]}&\ev{Dealloc\ \loc} \\
    \ev{Use\ \loc\ n}&\le_{\trace[_{\mssafe}]}&\ev{Dealloc\ \loc} \\
    \text{at most one }\ev{Dealloc\ \loc}&\text{in}&\trace[_{\mssafe}] \\
    \text{at most one }\ev{Alloc\ \loc\ n}&\text{in}&\trace[_{\mssafe}] \\
  \end{array}\right.\right\}
  $$
\end{definition}
Hereby, the notation $\event[_{1}]\le_{\trace}\event[_{2}]$ means that if $\event[_{1}]$ is in $\trace$ and if $\event[_{2}]$ is in $\trace$, then $\event[_{1}]$ appears before $\event[_{2}]$.

\subsubsection{Spatial Memory Safety}

\gls*{sms}~\cite{nagarakatte2009soft} is a safety property that prohibits out-of-bounds accesses.

\begin{definition}[\glsfirst*{sms}]\label{def:trace:smsdef}

  \noindent
  \[
  \smssafe:=\left\{\trace[_{\mssafe}] \left|\begin{array}{rcl}
      \text{If }\ev{Alloc\ \loc\ n}\le_{\trace_{\mssafe}}\ev{Use\ \loc\ m}, \text{ then }m<n
  \end{array}\right.\right\}
  \]
\end{definition}

\subsubsection{Memory Safety}

In spirit of earlier work~\cite{nagarakatte2009soft,nagarakatte2010cets,jim2002cyclone,necula2005ccured,michael2023mswasm}, full \gls*{ms} is the intersection of \Cref{def:trace:tmsdef,def:trace:smsdef}.

\begin{definition}[\glsfirst*{ms}]\label{def:trace:msdef}
  $
  \mssafe:=\tmssafe \cap \smssafe
  $
\end{definition}

Note that \Cref{def:trace:msdef} ignores data isolation, so there may still be memory-safety issues introduced by side-channels.

\subsection{A Trace Model for Memory Safety with Constant Time}\label{subsec:scct:tracemodel}

To express Constant Time, we extend the memory safety trace model with a {\em security tag} ($\securitytag{}$) that indicates whether events contains sensitive information ($\lock$) or not ($\unlock$).

\vspace{-.5em}
\[
  \begin{array}{rrcl}
    (\text{Base-Events}) & \preevent[_{\ctsafe}] &:=& \preevent[_{\mssafe}] \mid \ev{Branch\ }n \mid \ev{Binop\ }n\\
    (\text{Security Tags}) & \securitytag{} &:=& \lock \mid \unlock\\ 
    (\text{Events}) & \event[_{\ctsafe}] &:=& \preevent[_{\ctsafe}];\securitytag{} \mid \emptyevent \mid \lightning \\ 
  \end{array}
\]

For cryptographic code, there is a general guideline that secrets must not be visible on a trace~\cite{ctguidelines}, i.e., secrets should not be marked as $\unlock$.
In turn, an instruction whose timing is data-dependent must not have a secret as an operand.
Typical operations with data-dependent timing are branches and certain binary operations, such as division.%
\footnote{
	This is architecture-dependent, but division is an operation that serves as a classic example for a data-dependent timing instruction~\cite[p.~755]{arm-refman}.
}
Both operations are represented in the trace model by extending the set of base-events with branches ($\ev{Branch\ n}$) and binary operations ($\ev{Binop\ n}$).

\subsubsection{Strict Cryptographic Constant Time}

\gls*{cct} is a hypersafety property~\cite{barthe2018sec} and, thus, difficult to check with monitors.
This is because, intuitively, hypersafety properties can relate multiple execution traces with each other, but monitors work on a single execution.
It is a common trick to sidestep this issue by means of overapproximation: this section defines the property \gls*{scct}, a stricter variant of \gls*{cct} (inspired by earlier work~\cite{almeida2017jasmin}) that enforces the policy that no secret appears on a trace.
Programs that satisfy \gls*{scct} also satisfy \gls*{cct}, but programs that satisfy \gls*{cct} may not satisfy \gls*{scct}.

\begin{definition}[\glsfirst*{scct}]\label{def:trace:scctdef}

  \noindent\[
  \scctsafe:=\left\{\trace[_{\ctsafe}] \left|\begin{array}{l}
      \trace[_{\ctsafe}]=\hole{\cdot} \text{ or } \\\exists\trace[_{\ctsafe}'],\trace[_{\ctsafe}]=\preevent[_{\ctsafe}];\unlock\cdot\trace[_{\ctsafe}'] \wedge \trace[_{\ctsafe}']\in\scctsafe
    \end{array}\right.\right\}
  \]
\end{definition}

\gls*{scct} may appear overly strict, since it seems that secrets must not occur on a trace (since $\securitytag{}$ is forced to be $\unlock$). 
However, this is considered standard practice in terms of coding guidelines~\cite{ctguidelines}.
Moreover, programs that have been compiled with FaCT~\cite{cauligi2019fact} and run with a ``data independent timing mode''~\cite{arm-refman,intel-refman} enabled do not leak secrets (see \Cref{ex:lscct}). 

\subsubsection{\gls*{ms}, Strict Cryptographic Constant Time}\label{sec:msscct-rel}

The combination of \gls*{ms} and \gls*{scct} is the intersection of these properties, \gls*{msscct}.
However, \gls*{ms} uses a different trace model than \gls*{scct}, so intersecting them would trivially yield the empty set. 
To remedy this issue, we introduce $\sim_{\ctsafe}: \preevent[_{\ctsafe}] \times \preevent[_{\mssafe}] $, a cross-language trace relation (whose key cases are presented below), that we use to intuitively unify the trace model in which the two properties are expressed:
\[
  \typerulenolabel{mscctrel:drop:tag}{}{\preevent[_{\mssafe}]\sim_{\ctsafe}\preevent[_{\mssafe}];\unlock{}}
  \typerulenolabel{mscctrel:drop:crash}{}{\lightning\sim_{\ctsafe}\lightning}
\]
\[
  \typerulenolabel{mscctrel:drop:branch}{}{\emptyevent\sim_{\ctsafe}\ev{Branch n};\securitytag{}}
  \typerulenolabel{mscctrel:drop:binop}{}{\emptyevent\sim_{\ctsafe}\ev{Binop n};\securitytag{}}
\]

Essentially, $\sim_{\ctsafe}$ ignores both the new $\ev{Branch\ n}$ and $\ev{Binop\ n}$ base-events as it relates security-insensitive actions ($\unlock$) to their equivalent counterparts.
Thus, \gls*{ms} traces trivially satisfy \gls*{scct}.
The above relation is extended point-wise to traces, skipping the empty event $\emptyevent$ on either side, and it is now possible to define \gls*{msscct} using the universal image:

\begin{definition}[\gls*{ms} and \gls*{scct}]\label{def:trace:msscctdef}
  $
  \msscctsafe:=\mssafe\cap\sigma_{\sim_{\ctsafe}}\left(\scctsafe\right)
  $
\end{definition}

\subsection{Extending the Trace Model with Speculation}\label{subsec:msctss:tracemodel}

So far, the considered trace models do not let us express speculative execution attacks such as Spectre~\cite{kocher2019spectre}. 
For this, we extend the earlier trace model (see \Cref{subsec:scct:tracemodel}) so that the security tags ($\securitytag{}$) carry additional information about the kind of private data leakage, i.e., the type of speculative leak.
Moreover, we add base-events signalling the beginning of a speculative execution ($\ev{Spec}$), a barrier ($\ev{Barrier}$) that signals that any speculative execution may not go past it, as well as a rollback event ($\ev{Rlb}$), which signals that execution resumes to where speculation started.

\vspace{-1em}
{
\[
  \begin{array}{rrcl}
    (\text{Base-Events}) & \preevent[_{\mathghost}] &:=& \preevent[_{\ctsafe}] \\
    (\text{Spectre Variants}) & vX &:=& \operatorname{NONE} \mid \operatorname{PHT} \\
    (\text{Security Tags}) & \securitytag{} &:=& \lock_{vX} \mid \unlock\\ 
    (\text{Events}) & \event[_{\mathghost}] &:=& \preevent[_{\mathghost}];\securitytag{} \mid \emptyevent \mid \lightning \mid \ev{Spec} \mid \ev{Rlb} \mid \ev{Barrier}\\ 
  \end{array}
\]
}

Even though the considered Spectre variants are just SPECTRE-PHT~\cite{kocher2019spectre}, NONE just describes secret data as in \gls*{scct} (see \Cref{subsec:scct:tracemodel}), the trace model is general enough to allow for potential future extension with different variants~\cite{kocher2019spectre,maisuradze2018ret2spec,horn2019zero}.

\subsubsection{Speculative Safety}

\gls*{ss}~\cite{patrignani2021exorcising}, similar to \gls*{scct}, is a sound overapproximation of a variant of noninterference.

\begin{definition}[\glsfirst*{ss}]\label{def:trace:ss}
  \noindent

  \begin{nscenter}
  $
    \sssafe:=\left\{\trace[_{\mathghost}] \left|\begin{array}{l}
      \trace[_{\mathghost}]=\hole{\cdot} \text{ or } \exists\trace[_{\mathghost}'].\\
      \left(\trace[_{\mathghost}]=\preevent[_{\mathghost}];\unlock\cdot\trace[_{\mathghost}'] \text{or }\trace[_{\mathghost}]=\preevent[_{\mathghost}];\lock_{\text{NONE}}\cdot\trace[_{\mathghost}']\right)\\
      \text{and }\ \trace[_{\mathghost}']\in\sssafe
                                 \end{array}\right.\right\}
  $ 
  \end{nscenter}
\end{definition}
The technical setup so far leads to the above definition, where only locks annotated with $\text{SPECTRE-PHT}$ are disallowed to occur on the trace.
That way, programs attaining \gls*{ss} do not necessarily attain \gls*{scct}.

\subsubsection{Speculation Memory Safety}\label{sec:spec-ms-rel}

As before, we need to relate the different trace models with each other, so that the memory safety property without speculation can be lifted to speculation. 
To this end, let $\sim_{\mathghost}: \preevent[_{\mathghost}] \times \preevent[_{\ctsafe}]$ be a cross-language trace relation whose key cases are below.
The intuition is that \gls*{ss} is trivially satisfied in \gls*{scct}, since speculation is inexpressible there, which amounts to dropping events $\ev{Spec}$, $\ev{Rlb}$, or $\ev{Barrier}$, as well as all base events tagged with $\lock_{\text{PHT}}$. 

\[
  \typerulenolabel{specrel:drop:tag}{}{\preevent[_{\ctsafe}];\lock\sim_{\mathghost}\preevent[_{\ctsafe}];\lock_{\text{NONE}}}
  \typerulenolabel{specrel:drop:spectag}{}{\emptyevent\sim_{\mathghost}\preevent[_{\mathghost}];\lock_{\text{PHT}}}
\]
\[
  \typerulenolabel{specrel:drop:pub}{}{\preevent[_{\ctsafe}];\unlock\sim_{\mathghost}\preevent[_{\ctsafe}];\unlock}
  \typerulenolabel{specrel:drop:crash}{}{\lightning\sim_{\mathghost}\lightning}
\]
\[
  \typerulenolabel{specrel:drop:spec}{}{\emptyevent\sim_{\mathghost}\ev{Spec}}
  \typerulenolabel{specrel:drop:rlb}{}{\emptyevent\sim_{\mathghost}\ev{Rlb}}
  \typerulenolabel{specrel:drop:barrier}{}{\emptyevent\sim_{\mathghost}\ev{Barrier}}
\]

We conclude by defining the ultimate property of interest for secure compilers: \gls*{specms}.
\begin{definition}[\gls*{specms}]\label{def:trace:specmsdef}
  $
  \specmssafe := \msscctsafe\cap\sigma_{\sim_{\mathghost}}\left(\sssafe\right)
  $
\end{definition}

\section{Case Study: Language Formalisations\pages{2}}\label{sec:casestud:defs}

This section defines programming languages $\Ltms$, $\Ltrg$, $\Lms$, $\Lscct$, and $\Lspec$, all of which share common elements (presented in \Cref{subsec:cs:defs}).
$\Ltms$ is the only statically-typed language, and it exhibits the property that all well-typed programs are \gls*{tms} (\Cref{subsec:ltms}).
However, not all $\Ltms$ programs are \gls*{sms}.
That is, there are well-typed $\Ltms$ programs that perform an out-of-bounds access.
Language $\Ltrg$ is untyped and does not provide any guarantees with regards to \gls*{ms} (\Cref{subsec:lsms}).
$\Lms$ is exactly the same language as $\Ltrg$, but this paper still distinguishes the two for sake of readability (\Cref{subsec:lms}).
All three languages --- so $\Ltms$, $\Ltrg$, and $\Lms$ --- assume \gls*{scct} to hold, since this is -- in an ideal world -- what the programmer would expect, too: it is the job of the compiler to preserve and (potentially) enforce \gls*{scct} security~\cite{cauligi2019fact,nagarakatte2010cets,nagarakatte2009soft,akritidis2009baggy}.

Such consideration is also backed up by architecture providing a data (operand) independent timing mode, such as processors by Arm~\cite[p.~543]{arm-refman} and Intel~\cite[p.~80]{intel-refman}.
This kind of processor feature is modelled in language $\Lscct$ (\Cref{subsec:lscct}), where programs have access to a ``\gls*{cct}-mode'' and can change the leakage of emitted events according to the value of this mode (either $\obj{ON}$ or $\obj{OFF}$). 

Finally, modern processors also employ speculative execution to achieve speedups---and unfortunately generate Spectre attacks~\cite{kocher2019spectre}---and this is the extension of $\Lspec$ (\Cref{subsec:lspec}).
Thus, all previous languages trivially satisfy \gls*{ss}, since they do not support speculative execution at all.

\subsection{Shared Language Definitions}\label{subsec:cs:defs}

All presented programming languages share a common fragment which is partially presented here and in full detail in the technical report. 
{
  \renewcommand{\src}[1]{\mathsf{#1}}
 \small
\[
  \begin{array}{rrcl}
    \text{(Base-Events)} & \src{\preevent} &:=& \src{\ev{Alloc \loc\ n}} \mid \src{\ev{Dealloc \loc}} \mid \src{\ev{Get \loc\ n}} \mid \src{\ev{Set \loc\ n}} \\
    \text{(Control Tags)} & \src{\sandboxtag{}} &:=& \src{\ctx} \mid \src{\comp} \\
    \text{(Events)} & \src{\event} &:=& \src{\preevent;\sandboxtag{}} \mid \src{\emptyevent} \mid \src{\lightning} \\ 
    \text{(Values)} & \src{v} &:=& \src{n} \\
    \text{(Expressions)} & \src{\expr} &:=& \src{x} \mid \src{n} \mid \src{\lbinop{\expr[_1]}{\expr[_2]}} \mid \src{\lifz{\expr[_1]}{\expr[_2]}{\expr[_3]}} \\ 
                         &&&\mid \src{\llet{x}{\expr[_1]}{\expr[_2]}} \mid \src{\llet{x}{new\ \expr[_1]}{\expr[_2]}} \\
                         &&&\mid \src{\ldelete{x}} \mid \src{\lget{x}{\expr}} \mid \src{\lset{x}{\expr[_1]}{\expr[]}} \mid \src{stuck}\\
                         &&&\mid \src{\lcall{f}{\expr}} \mid \src{\lreturn{\expr}} \\
    \text{(Functions)} & \src{F} &:=& (\src{x};\src{\expr}) \\
    \text{(Libraries)} &&&\hspace{-3.8em} \src{\library_{\ctx}},\src{\library_{\comp}} : \src{Vars} \to \src{F} 
     \\
    \text{(Heaps)} & \src{H} &:=& \src{\hole{\cdot}} \mid \src{n},\src{H}
    \\
    \text{(Pointer Maps)} & \src{\Delta} && \text{ omitted for simplicity}
    \\
    \text{(Memory States)} & \src{\memstate} &:=& \left(\src{H^{\ctx}};\src{H^{\comp}};\src{\Delta}\right)\\
    \text{(Control States)} & \src{\cfstate} & & \text{ omitted for simplicity} 
    \\
    \text{(States)} &\src{\Omega} &:=& \left(\src{\cfstate};\src{\sandboxtag{}};\src{\memstate}\right)\\
  \end{array}
\]
}
{
  \renewcommand{\src}[1]{\mathsf{#1}}
The trace models of all languages are similar to those presented earlier (\Cref{sec:compprop}).
One technical detail is the addition of a control tag, indicating who is to blame for an emitted action: context ($\ctx$) or component ($\comp$).
This is a standard technique in secure compilation in order to rule out irrelevant context-level events. 
For example, a context immediately deallocating an allocated memory region twice trivially violates memory safety, but the main interest in secure compilation is to preserve component-level security, and thus component events.
Even though this tagging could be used for blame preservation~\cite{patrignani2023blame}, this is beyond the focus of this paper.
Another key difference is that memory accesses are now explicitly modelled as reads ($\src{\ev{Read\ \loc\ n}}$) and writes ($\src{\ev{Write\ \loc\ n}}$) instead of just uses.

All languages have at least numbers as values ($\src{v}$) and second class functions ($\src{F}$).
Functions are modelled as pairs containing the name of one argument and the body of the function.
Bodies are just ordinary expressions, which can be simple binary operations ($\src{\lbinop{\expr[_1]}{\expr[_2]}}$), conditionals ($\src{\lifz{\expr[_1]}{\expr[_2]}{\expr[_3]}}$), function calls ($\src{\lcall{f}{e}}$) and returns ($\src{\lreturn{\expr}}$), as well as C-like memory operations. 
Programs have sets of pre-determined functions called libraries and they are marked as being part of some component ($\src{\library_{\comp}}$) or context ($\src{\library_{\ctx}}$).

For the operational semantics, the runtime state ($\src{\Omega}$) is a triple consisting of a control-flow state ($\src{\cfstate}$), a control tag ($\src{\sandboxtag{}}$), and a memory state ($\src{\memstate}$). 
The latter carries information about pointers that are kept ``alive'' in pointer maps ($\src{\Delta}$), so that the semantics does not get stuck when encountering, e.g., a double-free.
The memory state also carries two heaps to model sandboxing between a context ($\src{H^{\ctx}}$) and a component ($\src{H^{\comp}}$).

}

\subsection{$\src{L_{\tmssafe}}$: A Temporal but Not Spatial Memory Safe Language}\label{subsec:ltms}

$\src{L_{\tmssafe}}$ is the only statically-typed language in this case study and restricts functions ($\src{F}$) to the typing signature $\src{\natt\to\natt}$. 
The type system of $\src{L_{\tmssafe}}$ is inspired by $L^{3}$~\cite{morrisett2005L3,scherer2018fabulous} and enforces that every well-typed $\src{L_{\tmssafe}}$ program satisfies \gls*{tms}.

\begin{theorem}[$\src{L_{\tmssafe}}$-programs are \gls*{tms}]\label{thm:wt:tms}
$\rsat{\src{\library_{\comp}}}{\tmssafe}$ \Coqed
\end{theorem}

\subsection{$\Ltrg$: A Memory-Unsafe Language}\label{subsec:lsms}

$\Ltrg$ extends the syntax presented earlier (\Cref{subsec:cs:defs}) with dynamic typechecks ($\trg{\lhast{\expr}{\type}}$), evaluating to $\trg{0}$ if the type matches with the shape of $\trg{\expr}$ and $\trg{1}$ otherwise.
Furthermore, the syntax of $\Ltrg$ is extended with a way to inspect whether a pointer is freed ($\trg{\lispoisoned{x}}$), evaluating to $\trg{0}$ if it is freed and to $\trg{1}$ otherwise.
Functions may receive arguments that are not $\trg{\natt}$, values are extended with tuples, and expressions are also extended with pair projections.

\[
  \begin{array}{rrcl}
    \text{(Values)} & \trg{v} &:=& \dots \mid \trg{\lpair{v_1}{v_2}} \\
    \text{(Expressions)} & \trg{\expr} &:=& \dots \mid \trg{\lhast{\expr}{\tau}} \mid \trg{\lispoisoned{x}}
  \end{array}
\]

No changes are done to the trace model, but $\Ltrg$ has no static typing, making double-free code patterns possible.

\subsection{$\Lms$: Another Memory-Unsafe Language}\label{subsec:lms}
$\Lms$ is exactly equal to $\Ltrg$ (\Cref{subsec:lsms}), but used to emphasize that this is code after applying $\cc{\Ltrg}{\Lms}$ (\Cref{subsec:cs:ms}).

\subsection{$\Lscct$: A Memory-Unsafe Language with a Constant Time Mode}\label{subsec:lscct}

$\Lscct$ extends $\Lms$ (\Cref{subsec:lms}) with a ,,constant-time mode''. 
The activation of the mode can be checked ($\obj{\lrddoit{x}{\expr}}$), changed ($\obj{\lwrdoit{D}}$), and is stored in program states ($\obj{\Omega}$).
At the beginning of program execution of $\Lscct$ programs, the mode is turned off ($\obj{OFF}$).
If the mode is enabled (i.e., set to $\obj{ON}$), the intuition is that no secrets are leaked. 
This models the real-world\footnote{As present in Intel~\cite[p.80]{intel-refman} and ARM~\cite[p.~543]{arm-refman} processors.} data-independent timing mode as well as the result of compiling a program with FaCT~\cite{cauligi2019fact}.
For example, FaCT rewrite code that branches to use a constant-time selection primitive.
To not obfuscate our formalisation unnecessarily by duplicating syntax, we simply added the ,,constant-time mode'' to $\Lscct$.

\[
  \begin{array}{rrcl}
    \text{(Mode Values)} & \obj{D} &:=& \obj{ON} \mid \obj{OFF} \\
    \text{(Security Tags)} & \obj{\securitytag{}} &:=& \obj{\lock} \mid \obj{\unlock} \\
    \text{(Base-Events)} & \obj{\preevent} &:=& \dots \mid \obj{iGet\ \loc\ v} \mid \obj{iSet\ \loc\ v} \\
                       &&& \mid\ \obj{Binop\ n} \mid \obj{Branch\ n} \\
    \text{(Events)} & \obj{\event} &:=& \obj{\preevent;\sandboxtag{};\securitytag{}} \mid \obj{\emptyevent} \mid \obj{\lightning} \\
    \text{(Expressions)} & \obj{\expr} &:=& \dots \mid \obj{\lrddoit{x}{\expr}} \mid \obj{\lwrdoit{D}} \\
                         &&&\mid\ \obj{\llet{x^{\securitytag{}}}{\expr[_1]}{\expr[_2]}}\\
    \text{(States)} & \obj{\Omega} & := & \left(\obj{\cfstate};\obj{\sandboxtag{}};\obj{D};\obj{\memstate}\right)\\
  \end{array}
\]
\begin{center}
\newcommand{\expreval}[5]{{#1}\triangleright\xspace {#2}\xrightarrow{#5}\ {#3}\triangleright\xspace {#4}\xspace}
\newcommand{\exprevalo}[5]{\expreval{\obj{#1}}{\obj{#2}}{\obj{#3}}{\obj{#4}}{\obj{#5}}}
  \typerule{$e-\obj{wrdoit}-\text{off}$}{
  }{
    \pstepto{\rtt{\obj{\cfstate;\sandboxtag{};D;\memstate}}{\obj{\lwrdoit{OFF}}}}{\rtt{\obj{\cfstate;\sandboxtag{};OFF;\memstate}}{\obj{0^{\securitytag{}}}}}{\obj{\emptyevent}}
  }{e-wrdoit-off}

  \typerule{$e-\obj{get}-\in-$noleak}{
    \obj{\memstate}=\obj{H^{\ctx};H^{\comp};\Delta_1,x\mapsto(\loc;\sandboxtag{};\poison),\Delta_2} &
    \obj{\loc}+\obj{n}\in\text{dom }\obj{H^{\sandboxtag{}}} \\
    \obj{\securitytag{''}}=\obj{\securitytag{}\sqcap\securitytag{'}} &
    \obj{\event} = \obj{{iGet}\ \loc\ n;\sandboxtag{};\securitytag{''}}
  }{
    \exprevalo{
    	\cfstate;\sandboxtag{'};ON;\memstate
    }{
      	x^{\securitytag[n^{\securitytag{'}}] }
    }{
    	\cfstate;\sandboxtag{'};ON;\memstate
    }{
    	(H^{\sandboxtag{}}(\loc+n))^{\securitytag{''}}
    }{
      \event
    }
  }{to-e-get-in-noleak}
\end{center}

The language also adds user annotations $\obj{\securitytag{}}$ for the secrecy of variables, which can be either private (high secrecy) $\obj{\lock}$ or public (low secrecy) $\obj{\unlock}$.
Security tags $\obj{\securitytag{}}$ are arranged in the usual secrecy lattice~\cite{zdancewicphd}, where $\obj{\unlock}\sqsubseteq\obj{\lock}$.

Memory accesses to secret data need to be present to reason about memory safety, even when execution is in constant-time mode, e.g, \Cref{tr:to-e-get-in-noleak}.
$\Lscct$ extends base-events with $\obj{\ev{iGet}\ \loc\ \valueexpr}$ and $\obj{\ev{iSet}\ \loc\ \valueexpr}$ (for data $\obj{i}$ndependent get and set) to prevent secrets from leaking but still enable reasoning about memory safety. 
Due to the technical setup, the rule needs to check if the access is in bounds ($\obj{\loc}+\obj{n}\in\obj{H^t}$) and update the secrecy tag ($\obj{\securitytag{''}}=\obj{\securitytag{}\sqcap\securitytag{'}}$) with the least upper bound of the tags, according to the aforementioned lattice.
The precise information carried by the event, e.g., location ($\obj{\loc}$) is taken from the pointer map, which carries information irrelevant to this rule ($\obj{\poison}$).

Base-events include $\obj{\ev{Branch}\ n}$ and $\obj{\ev{Binop}\ n}$ that are emitted when evaluating a branch or certain binary expressions, such as division, respectively, whenever the constant-time mode is inactive.
Events are extended with a security-tag ($\obj{\securitytag{}}$) to signal the secrecy of the involved data.

The evaluation steps are amended to propagate the security-tag annotations $\obj{\securitytag{}}$.
When the constant-time mode is inactive, base-events $\obj{\ev{Branch}\ n}$ and $\obj{\ev{Binop}\ n}$ are emitted for conditionals and binary operations, respectively.
Otherwise, just like in the semantics of the earlier languages, $\obj{\emptyevent}$ is emitted for binary and branching operations.

\Cref{ex:lscct} illustrates the differences between $\Lscct$ and other languages.

\begin{exampleenv}[$\Lscct$ with and without constant-time mode]\label{ex:lscct}
  Consider again \Cref{ex:strncpy}, with a context copying the string \texttt{Hello World}, where everything is marked with a high security tag: $\lock$.
  The top half of \Cref{fig:ex-cct} (titled $\Lscct$), describes the execution trace of the program, while the bottom side of the table (titled ``Specification''), describes the related specification trace (\Cref{subsec:msctss:tracemodel}).
  Read in parallel from top to bottom, the figure shows parts of the execution trace. 
  In each half, the left column (\textit{Active}) has constant-time mode $\obj{ON}$ and the right one (\textit{Inactive}) has it $\obj{OFF}$.
  
  When the constant-time mode is off, the execution yields events in similar fashion to before (\Cref{subsec:ltms,subsec:lsms,subsec:lms}).
  But, if it is turned on, then the branching event \colorbox{yellow}{does not fire} anymore and both reading and writing to memory is related to a specification trace with \colorbox{lightgray}{no exposed secrets}.
\end{exampleenv}
\begin{figure}[!htb]
	\vspace{-1em}\small
  $$
  \begin{array}{ccc}
    & \Lscct &        \\
    \text{Active} & \mid & \text{Inactive} \\\hline
    \obj{\ev{Alloc}\ \loc_{x}\ 12;\comp;\unlock} & \mid & \obj{\ev{Alloc}\ \loc_{x}\ 12;\comp;\unlock} \\
    \obj{\ev{Alloc}\ \loc_{y}\ 12;\comp;\unlock} & \mid & \obj{\ev{Alloc}\ \loc_{y}\ 12;\comp;\unlock} \\
    \obj{{\ev{iGet}}\ \loc_{x}\ 0;\comp;}\text{\colorbox{lightgray}{$\obj{\lock}$}} & \mid & \obj{\ev{Get}\ \loc_{x}\ 0;\comp;\lock} \\
    \text{\colorbox{yellow}{$\obj{\emptyevent}$}} & \mid & \obj{\ev{Branch}\ 0;\comp;\lock} \\
    \obj{{\ev{iGet}}\ \loc_{x}\ 0;\comp;}\text{\colorbox{lightgray}{$\obj{\lock}$}} & \mid & \obj{\ev{Get}\ \loc_{x}\ 0;\comp;\lock} \\
    \obj{{\ev{iSet}}\ \loc_{y}\ 0\ \mathtt{'H'};\comp;}\text{\colorbox{lightgray}{$\obj{\lock}$}} & \mid & \obj{\ev{Set}\ \loc_{y}\ 0\ \mathtt{'H'};\comp;\lock} \\
    \obj{{\ev{iGet}}\ \loc_{x}\ 1;\comp;}\text{\colorbox{lightgray}{$\obj{\lock}$}} & \mid & \obj{\ev{Get}\ \loc_{x}\ 1;\comp;\lock} \\
    \text{\colorbox{yellow}{$\obj{\emptyevent}$}} & \mid & \obj{\ev{Branch}\ 0;\comp;\lock} \\
    \vdots & \mid & \vdots \\
    \obj{{\ev{iGet}}\ \loc_{x}\ 12;\comp;}\text{\colorbox{lightgray}{$\obj{\lock}$}} & \mid & \obj{\ev{Get}\ \loc_{x}\ 12;\comp;\lock} \\
    \text{\colorbox{yellow}{$\obj{\emptyevent}$}}& \mid & \obj{\ev{Branch}\ 1;\comp;\lock} \\
    \obj{\ev{Dealloc}\ \loc_{y};\comp;\unlock} & \mid & \obj{\ev{Dealloc}\ \loc_{y};\comp;\lock} \\
    \obj{{\ev{iGet}}\ \loc_{y}\ 6;\comp;}\text{\colorbox{lightgray}{$\obj{\lock}$}} & \mid & \obj{\ev{Get}\ \loc_{y}\ 6;\comp;\lock} \\
  \end{array}
  $$
  $$
  \begin{array}{ccc}
          & \text{Specification} & \\
  \text{Active} & \mid & \text{Inactive} \\\hline
     \ev{Alloc}\ \loc_{x}\ 12;\unlock & \mid & \ev{Alloc}\ \loc_{x}\ 12;\unlock\\
     \ev{Alloc}\ \loc_{y}\ 12;\unlock & \mid & \ev{Alloc}\ \loc_{x}\ 12;\unlock\\
     \ev{Use}\ \loc_{x}\ 0;\text{\colorbox{lightgray}{$\unlock$}} & \mid & \ev{Use}\ \loc_{x}\ 0;\lock\\
     \emptyevent & \mid & \ev{Branch}\ 0;\lock\\
     \ev{Use}\ \loc_{x}\ 0;\text{\colorbox{lightgray}{$\unlock$}} & \mid & \ev{Use}\ \loc_{x}\ 0;\lock\\
     \ev{Use}\ \loc_{y}\ 0;\text{\colorbox{lightgray}{$\unlock$}} & \mid & \ev{Use}\ \loc_{y}\ 0;\lock\\
     \ev{Use}\ \loc_{x}\ 1;\text{\colorbox{lightgray}{$\unlock$}} & \mid & \ev{Use}\ \loc_{x}\ 1;\lock\\
     \emptyevent & \mid & \ev{Branch}\ 0;\lock\\
     \vdots & \mid & \vdots \\
      \ev{Use}\ \loc_{x}\ 12;\text{\colorbox{lightgray}{$\unlock$}} & \mid & \ev{Use}\ \loc_{x}\ 12;\lock\\
      \emptyevent & \mid & \ev{Branch}\ 1;\lock\\
      \ev{Dealloc}\ \loc_{y};\unlock & \mid & \ev{Dealloc}\ \loc_{y};\unlock\\
      \ev{Use}\ \loc_{y}\ 6;\text{\colorbox{lightgray}{$\unlock$}} & \mid & \ev{Use}\ \loc_{y}\ 6;\lock\\
  \end{array}
  $$
  \caption{Traces for \Cref{ex:lscct}.}
  \label{fig:ex-cct}
\end{figure}

\subsection{$\Lspec$: A Memory-Unsafe Language with Speculation}\label{subsec:lspec}

$\Lspec$ extends $\Lscct$ (\Cref{subsec:lscct}) with speculative dynamic semantics that is inspired by existing work~\cite{guarnieri2018spectector,fabian2022automatic}.
After branching, speculation starts by pushing the current configuration into the speculation state ($\ird{S}$), and run subsequent code in a predetermined window (whose size is $\omega$) until a rollback of operational state is performed. 
The window is set inside the speculation state.
Within such windows, data may leak, this is marked as high $\ird{\lock}$ and with an annotation that indicates the respective speculative execution variant. 
In this paper, we just consider SPECTRE-PHT~\cite{kocher2019spectre}, whose starting point is \Cref{tr:e-ifz-true-spec}. %
Additionally, $\ird{\lock}$ may also carry a $\ird{NONE}$ annotation, to signal that this is a leak irrespective of speculation, i.e., as defined earlier (\Cref{subsec:lscct}).
The semantics is kept general enough to allow for future extension to support different variants~\cite{kocher2019spectre,maisuradze2018ret2spec,horn2019zero}.

In terms of language features, $\Lspec$ includes a new barrier operation $\ird{\lbarrier}$ that blocks speculative execution. 
To facilitate speculative execution in a non-assembly-like language, a stack of operational state is used and speculation is active if that stack has a size greater than one.
This is exploited in \Cref{tr:e-spec-eat} to consume any leftover speculation.

\vspace{-.5em}
\[
  \begin{array}{rrcl}
    \text{(Base-Events)} & \ird{\preevent} &:=& \dots \mid \ird{Spec} \mid \ird{Rlb} \mid \ird{Barrier}\\
    \text{(Expressions)} & \ird{\expr} &:=& \dots \mid \ird{\lbarrier} \\
    \text{(Speculation State)} & \ird{S} &:=& \ird{\Omega\triangleright\expr} \mid \ird{S},\left(\ird{\Omega};\ird{n};\ird{\expr}\right)
  \end{array}
\]

\begin{center}
\newcommand{\expreval}[5]{{#1}\triangleright\xspace {#2}\xrightarrow{#5}\ {#3}\triangleright\xspace {#4}\xspace}
\newcommand{\exprevald}[5]{\expreval{\ird{#1}}{\ird{#2}}{\ird{#3}}{\ird{#4}}{\ird{#5}}}

  \typerule{$e-\ird{ifz}-\ird{true}-\text{spec}$}{
    \ird{S} = \rtt{(\ird{\cfstate['];\sandboxtag{};D;\memstate});\ird{S}}{\ird{\expr[_1]}},({\ird{\cfstate;\sandboxtag{};D;\memstate}};\omega;{\ird{\expr[_2]}})
  }{
    \pstepto{\rtt{\ird{\cfstate;\sandboxtag{};D;\memstate}}{\ird{\lifz{0^{\securitytag{}}}{\expr[_1]}{\expr[_2]}}}}{\ird{S}}{\ird{{Branch}\ 0;\sandboxtag{};\securitytag{}}\cdot\ird{{Spec};\sandboxtag{};\securitytag{}}}
  }{e-ifz-true-spec}
  \typerule{$e-\ird{spec}-\text{eat}$}{
    \ird{n} > 0 &
    \pstepto{\rtt{\ird{\Omega}}{\ird{\expr}}}{\rtt{\ird{\Omega'}}{\ird{\expr[']}}}{\ird{\event}}
  }{
    \ghoststepto{\ird{S},(\ird{\Omega};\ird{n};\ird{\expr})}{\ird{S},(\ird{\Omega'};\ird{n}-1;\ird{\expr'})}{\ird{\event}}
  }{e-spec-eat}
  \typerule{$e-\ird{spec}-\text{eaten}$}{
  }{
    \ghoststepto{\ird{S},(\ird{\Omega};\ird{0};\ird{\expr})}{\ird{S}}{\ird{Rlb;\Omega.\sandboxtag{};\unlock}}
  }{e-spec-eaten}
  \typerule{$e-\ird{spec}-\text{barrier}$}{
  }{
    \ghoststepto{\ird{S},(\ird{\Omega};\ird{n};\ird{\lbarrier})}{\ird{S},(\ird{\Omega};\ird{0};\ird{0})}{\ird{Barrier;\Omega.\sandboxtag{};\unlock}}
  }{e-spec-barrier}
\end{center}

The trace model is extended with $\ird{{Spec}}$, $\ird{{Rlb}}$, and $\ird{{Barrier}}$ events that signal the respective operational action. 
A $\ird{\lbarrier}$ prevents any execution whatsoever besides $\ird{{Rlb}}$ when run in speculative execution mode and does nothing when run in normal mode, e.g., \Cref{tr:e-spec-barrier}.%
\footnote{In the rule, the notation $\cfstate' = \cfstate[.\text{win} = 0]$ means that $\cfstate'$ is a copy of $\cfstate$ up to field $\text{win}$, which is set to $0$.}
Lastly, $\ird{Rlb}$ is emitted when the speculation window is zero and the operational state is rolled back (see \Cref{tr:e-spec-eaten}).

\section{Case Study: Composing Secure Compiler Passes and Optimisations \pages{7}}\label{sec:casestud:rtp}
\begin{figure*}[!h]
  \centering
  \begin{tikzpicture}
    \node (S) {$\src{L_{\tmssafe}}$};
    \node[right=2.0 of S] (T) {$\trg{L}$};
    \node[right=2.0 of T] (M) {$\irl{L_{\mssafe}}$};
    \node[below right=1.5 and 1.0 of M] (D0) {$\irl{L_{\mssafe}}$};
    \node[above right=1.5 and 1.0 of M] (C0) {$\irl{L_{\mssafe}}$};
    \node[right=3.0 of M] (E) {$\irl{L_{\mssafe}}$};
    \node[right=2.0 of E] (O) {$\obj{L_{\scctsafe}}$};
    \node[right=2.0 of O] (Z) {$\ird{L_{\mathghost}}$};

    \draw[->] (S) to[sloped] node[align=center] (tmsedge) {\gls*{tms}\\ \Cref{thm:cca:rtp:tms}} (T);
    \draw[->] (T) to[sloped] node[align=center] {\gls*{sms}\\ \Cref{thm:ccb:rtp:sms}} (M);
    \draw[->] (M) to[sloped] node[align=center] {\gls*{dce}\\ \Cref{thm:ccdce:rtp:ms}} (D0);
    \draw[->] (M) to[sloped] node[align=center] {\gls*{cf}\\ \Cref{thm:cccf:rtp:ms}} (C0);
    \draw[->] (D0) to[sloped] node[align=center] {\gls*{cf}\\ \Cref{thm:cccf:rtp:ms}} (E);
    \draw[->] (C0) to[sloped] node[align=center] {\gls*{dce}\\ \Cref{thm:ccdce:rtp:ms}} (E);
    \draw[->] (E) to[sloped] node[align=center] {\gls*{scct}\\ \Cref{thm:ccscct:rtp:scct}} (O);
    \draw[->] (O) to[sloped] node[align=center] {\gls*{ss}\\ \Cref{thm:ccspec:rtp:spec}} (Z);

    \node (sectms) at ($(S)+(1.25,2.5)$) {{\Cref{subsec:cs:tms}}};
    \node (secsms) at ($(T)+(1.25,2.5)$) {{\Cref{subsec:cs:ms}}};
    \node (secopts) at ($(M)+(2.0,2.5)$) {{\Cref{subsec:cs:opts}}};
    \node (secscct) at ($(E)+(1.5,2.5)$) {{\Cref{subsec:cs:scct}}};
    \node (secss) at ($(O)+(1.5,2.5)$) {{\Cref{subsec:cs:ss}}};

    \draw[thick,dotted,->] ($(S) + (0.2,-0.27)$) to ($(S) - (-0.2,1)$) to node[align=center] {\gls*{ms}\\ \Cref{corr:ccab:rtp:ms}} ($(M) - (0.0,1)$) to (M);

    \draw[thick,dotted,->] (S) to ($(S) - (0.0,2.5)$) to node[align=center,pos=.65] {\gls*{specms}\\ \Cref{thm:ccall:rtp:specms}} ($(Z) - (0.0,2.5)$) to (Z);

    \draw[thick,dotted,->] (M) to[bend right=0,sloped] node[align=center] {\gls*{ms}\\ \Cref{thm:cccfccdce:rtp:ms}} (E);

    \draw[dashed] ($(S)-(0.0,-0.5)$) -- ($(S)-(0.0,-2.5)$);
    \draw[dashed] ($(T)-(0.0,-0.5)$) -- ($(T)-(0.0,-2.5)$);
    \draw[dashed] ($(M)-(0.0,-0.5)$) -- ($(M)-(0.0,-2.5)$);
    \draw[dashed] ($(E)-(0.0,-0.5)$) -- ($(E)-(0.0,-2.5)$);
    \draw[dashed] ($(O)-(0.0,-0.5)$) -- ($(O)-(0.0,-2.5)$);
    \draw[dashed] ($(Z)-(0.0,-0.5)$) -- ($(Z)-(0.0,-2.5)$);
  \end{tikzpicture}
  \vspace{-1em}
  \caption{Visualisation of the optimising compilation pipeline that preserves \gls*{specms}. %
    Vertices in the graph are the programming languages from earlier sections (\Cref{sec:casestud:defs}). %
    Full edges are secure compilers passes.
    Dotted edges are composition of passes and use the presented framework (\Cref{sec:sequential}) to indicate the property they preserve. %
    The dashed lines partition the graph into the sections where the respective theorems are presented.
  }\label{fig:pipeline}
\end{figure*}
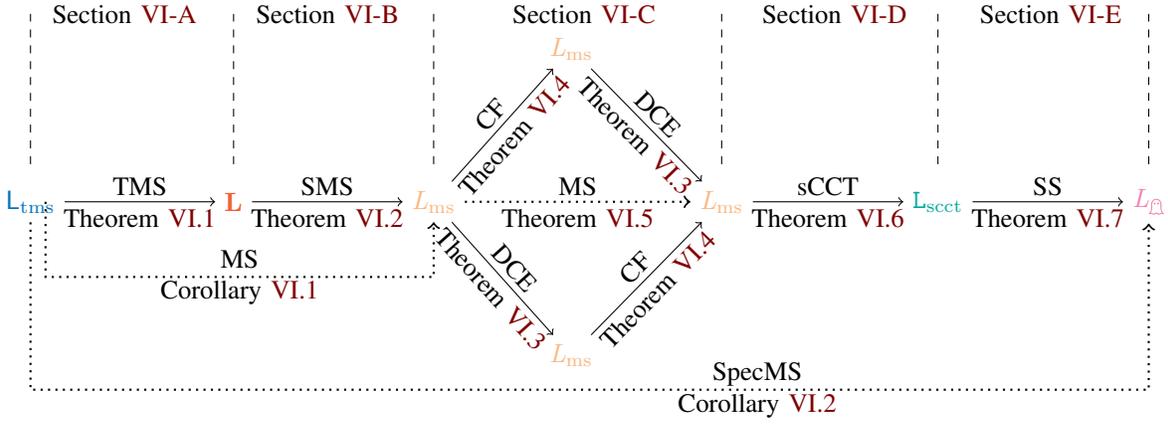

This section defines several secure compilers, each of which robustly preserves a different property of interest as depicted in \Cref{fig:pipeline}.
The section demonstrates the power of the framework (\Cref{sec:sequential}) by composing these compilers for a secure and optimising compilation chain that robustly preserves \gls*{specms}.
The first step in this chain is the compiler from $\src{L_{\tmssafe}}$ to $\trg{L}$ that robustly preserves just \gls*{tms} (\Cref{thm:cca:rtp:tms}).
From here, another pass from $\trg{L}$ to $\irl{L_{\mssafe}}$ ensures that no out-of-bounds accesses can happen and, thus, programs at this point attain \gls*{sms} (\Cref{thm:ccb:rtp:sms}).
Since these properties compose into \gls*{ms}, composing these passes yields a compiler that robustly preserves \gls*{ms} (\Cref{corr:ccab:rtp:ms}).
Then, the section presents two optimisation passes, namely \gls*{cf} and \gls*{dce}, each of which robustly preserves \gls*{ms} (\Cref{thm:ccdce:rtp:ms,thm:cccf:rtp:ms}).
These passes can be freely ordered in the compilation chain without compromising memory safety (\Cref{thm:cccfccdce:rtp:ms}).
The next step in the chain ensures that code stays \gls*{scct} (\Cref{thm:ccscct:rtp:scct}) when compiled from $\Lms$ to $\Lscct$, which is done by switching on a constant-tame mode for the computation.
Lastly, by introducing barriers immediately after branches, speculative leaks via SPECTRE-PHT are prevented when compiling $\Lscct$ to $\Lspec$.
The final result is that the whole compilation chain robustly preserves \gls*{specms} (\Cref{thm:ccall:rtp:specms}).

\subsection{Robust Temporal Memory Safety Preservation}\label{subsec:cs:tms}

The secure compiler from $\Ltms$ to $\Ltrg$ needs to ensure that when execution switches from context to component, the type signatures are respected.
Thus, the compiler inserts the following dynamic typechecks before entering the body of a component-defined function (anything elided is a trivial identity function from source to target):

\vspace{-1em}
{
\begin{gather}
  \begin{align*}%
    \cca(\src{\lget{x}{\expr}}) = &\ \lget[\trg]{\trg{x}}{\left[\cca(\src{\expr})\right]} \\
    \cca(\src{\ldelete{x}}) = &\ \ldelete[\trg]{\left[\cca(\src{x})\right]} 
    \\
    \cca(\src{\lfunction{g}{x:\natt\to\type[_{e}]}{\expr}})  =&
  \end{align*}
  \\
  \begin{align*}
\lfunction[\trg]{\trg{g}}{\trg{x}}{\lifz[\trg]{\trg{\lhast{x}{\natt}}}{
                                                            \left[\cca(\src{\expr})\right] %
                                                                                                 }{\labort[\trg]}}
  \end{align*}
\end{gather}
}

\noindent Since $\trg{L}$ has no static typing, an attacker $\trg{\library_{\ctx}}$ can invoke a component function accepting a $\src{\natt}$ with $\trg{\lpair{17}{29}}$.
With the dynamic check, the compiler ensures that execution aborts in such cases.

Compiling the \texttt{strncpy} function from \Cref{sec:introduction} with $\cca$, the compiler would in this case ensure that the arguments that are evaluated in the compiled \texttt{strncpy} are valid.

$\cca$ is robustly preserving (\Cref{def:rtp}) \gls*{tms}:
\begin{theorem}[$\cca$ secure w.r.t. \gls*{tms}]\label{thm:cca:rtp:tms}
    $\rtp{\cca}{\tmssafe}$ %
\end{theorem}

\subsection{Robust Spatial Memory Safety Preservation}\label{subsec:cs:ms}

The spatial memory safety preserving compiler from $\Ltrg$ to $\Lms$ only inserts bounds-checks whenever reading from or writing to memory in order to enforce \gls*{sms}.
These bounds checks need the bounds information, which the compiler keeps around by introducing a fresh identifier $\irl{x_{SIZE}}$ for each allocation that binds $\irl{x}$.
Then, it is simply a matter of referring to that variable and ensuring that memory accesses are in the interval $[\irl{0},\irl{x_{SIZE}})$.
When the check fails, the code aborts.

\begin{nscenter}
  $$
  \begin{array}{rcl}
    \ccb(\trg{\lnew{x}{\expr[_{1}]}{\expr[_{2}]}}) & = 
                                                   & \llet[\irl]{\irl{x_{SIZE}}}{\ccb(\trg{\expr[_{1}]})}{}
    		\\&&
    		\lnew[\irl]{\irl{x}}{\irl{x_{SIZE}}}{\ccb(\trg{\expr[_{2}]})}
    		 \\
  \ccb(\trg{\lget{x}{\expr}}) & = 
                              & \llet[\irl]{\irl{x_{n}}}{\ccb(\trg{\expr})}{}
  	\\&&
  \lifz[\irl]{\irl{0\le x_{n}<x_{SIZE}}}{\\&&\irl{\lget{x}{x_{n}}}}{}
  		\irl{\labort}
  	  \\
  \ccb(\trg{\lset{x}{\expr[_{1}]}{\expr[_{2}]}}) & = 
                                                 & \llet[\irl]{\irl{x_{n}}}{\ccb(\trg{\expr[_{1}]})}{}
  		\\&&
  \lifz[\irl]{\irl{0\le x_{n}<x_{SIZE}}}{\\&&\lset[\irl]{\irl{x}}{\irl{x_{n}}}{}
  		\ccb(\trg{\expr[_{2}]})
  		}{\irl{\labort}} 
  \end{array}
  $$
\end{nscenter}

\begin{exampleenv}[Instrumented \texttt{strncpy}]
  Consider again \texttt{strncpy}, but instrumented for \gls*{sms}:
    \begin{lstlisting}[language=c,basicstyle=\ttfamily, morekeywords={size_t}]
void strncpy(size_t n, size_t dst_size, char *dst,
             size_t src_size, char *src) {
  for(size_t i = 0; i < src_size
      && src[i] != '\0' && i < n; ++i) {
    if(i < src_size && i < dst_size) {
      dst[i] = src[i];
    }
  }
}
    \end{lstlisting}
Consider context \texttt{strncpy(2, x, y)}, where \texttt{x} and \texttt{y} are pointers to valid regions of memory with allocated space for exactly two cells and do not contain the null-terminating character \texttt{'\textbackslash 0'}.
Without the \gls*{sms} pass, the event $\ev{Use}\ \loc_{x}\ 2;\comp;\unlock$ would appear on the trace, but that indicates an out-of-bounds access! 
Fortunately, with \gls*{sms} mitigation in place, that event does not appear during execution, since the bounds check prevents the condition \texttt{src[i] != '\textbackslash 0'} from executing.
\end{exampleenv}

Contrary to the previous compiler, $\ccb$ may change the trace of the original $\Ltrg$ program: if there is a memory access, it needs to be protected with a bounds check.
The corresponding relation $\sim^{\Ltrg}_{\Lms} : \trg{\trace}\times\irl{\trace}$ that describes this semantic effect of the compiler is defined partially below.
We omit action $\trg{Set}$, which is treated analogously, and any other event, which is related to its cross-language equivalent.

{
\[
  \typerulenolabel{xrel:sms:read}{\trg{n}\text{ in bounds}}{\trg{Get\ \loc\ n;\comp}\sim^{\Ltrg}_{\Lms}\irl{Get\ \loc\ n;\comp}}
\]
}

For simplicity, we elide the environment that this relation carries around in order to bind each location to its metadata (such as its size), and resolve the ``$\trg{n}$ in/out of bounds'' premise.
We can now prove that compiler $\ccb$ robustly preserves \gls*{sms}.
\begin{theorem}[$\ccb$ secure w.r.t. \gls*{sms}]\label{thm:ccb:rtp:sms}
  $\rtpsim{\ccb}{\smssafe}{\sim^{\Ltrg}_{\Lms}}$ %
\end{theorem}

At this point we can compose $\ccb$ with the previous compiler ($\cca$), but in order to do so, we need a trace relation from $\Ltms$ to $\Lms$.
We can obtain this relation by composing the trace relation we just defined ($\sim^{\Ltrg}_{\Lms}$) with the one used by the previous compiler: $\sim^{\Ltms}_{\Ltrg} : \src{\trace}\times\trg{\trace}$.
The latter has not been previously defined (nor has it been used in the related theorem) because that is just an equality relation, since the trace models of $\Ltms$ and of $\Ltrg$ are the same.
Thus, we formally define $\sim^{\Ltms}_{\Lms}: \src{\trace} \times \irl{\trace}$ as the following composition: $\sim^{\Ltms}_{\Ltrg}\bullet\sim^{\Ltrg}_{\Lms}$.
With this relation, \Cref{corr:ccab:rtp:ms} states that the composition of $\cca$ and $\ccb$ is secure w.r.t. \gls*{ms} and it follows from \Cref{thm:cca:rtp:tms,thm:ccb:rtp:sms} using \Cref{thm:rtpsim:sig}.
\begin{corollary}[$\cca\circ\ccb$ secure w.r.t. \gls*{ms}]\label{corr:ccab:rtp:ms}
  $\;$ 

  \begin{nscenter}
    $\rtpsim{\cca\circ\ccb}{\mssafe}{\sim^{\Ltms}_{\Lms}}$ %
  \end{nscenter}
\end{corollary}
This proof requires another precondition besides \Cref{thm:ccb:rtp:sms,thm:cca:rtp:tms}: $\sim^{\Ltms}_{\Ltrg}$ needs to be well-formed with respect to $\smssafe$.
This follows trivially since $\sim^{\Ltms}_{\Ltrg}$ is an equality. 
\begin{lemma}[$\sim^{\Ltms}_{\Ltrg}$ well-formed w.r.t. $\smssafe$]\label{lem:wf:ltmsltrg}
$\;$ 

  \begin{nscenter}
  $\wfcsig{\sim^{\Ltms}_{\Ltrg}}{\smssafe}$
  \end{nscenter}
\end{lemma}

\subsection{Optimising Compilers}\label{subsec:cs:opts} 

This section defines two optimising compiler passes from $\Lms$ to $\Lms$ which perform \gls*{dce} and \gls*{cf}, respectively.
The \gls*{dce} pass applies a na\"ive rewrite rule on conditionals.
The \gls*{cf} pass relies on an auxiliary function \texttt{mix} that uses a substitutions accumulator $\irl{\substlist}$ in order to rewrite constant binary operations, e.g., $\irl{{17}-1}$ to $\irl{16}$, and replace variables that are assigned to constants, e.g., $\irl{\llet{x}{7}{x}}$ to $\irl{7}$.

\vspace{-1em}
\begin{gather*}
  \begin{align*}
    \ccdce(\irl{\lifz{true}{\expr[_{1}]}{\expr[_{2}]}}) & = \ccdce(\irl{\expr[_{1}]}) &\\
    \ccdce(\irl{\lifz{false}{\expr[_{1}]}{\expr[_{2}]}}) & = \ccdce(\irl{\expr[_{2}]}) &
  \end{align*}
  \\
  \begin{align*}
    \cccf(\irl{\expr}) & = \partialeval{\irl{\expr}}{\irl{\hole{\cdot}}} &
  \end{align*}
  \\[0.125cm]
  \begin{align*}
   \partialeval{\irl{x}}{\irl{\substlist}} & = \irl{n} 
   	\qquad\qquad \text{if } \irl{\subst{n}{x}}\in\irl{\substlist} \\
   \partialeval{\irl{x}}{\irl{\substlist}} & = \irl{x} 
   \qquad\qquad \text{if } \irl{\subst{n}{x}}\notin\irl{\substlist} \\
   \partialeval{\irl{\lbinop{n}{m}}}{\irl{\substlist}} & = \irl{k} 
   \qquad\qquad \text{if } \lbinop{\irl{n}}{\irl{m}}=k \\
   \partialeval{\irl{\llet{x}{n}{\expr}}}{\irl{\substlist}} & = \partialeval{\irl{\expr}}{\irl{\subst{n}{x}\cdot\substlist}} 
\end{align*}
\end{gather*}

Note that both passes have no effect on the resulting trace of a program, up to $\emptyevent$-steps. 
Because of this, both passes have equality as corresponding cross language trace relation. 
Moreover, it is straightforward to prove both passes as secure (\Cref{def:rtp}) w.r.t. \gls*{ms}. 

\begin{theorem}[$\ccdce$ secure w.r.t. \gls*{ms}]\label{thm:ccdce:rtp:ms}
  $\rtp{\ccdce}{\mssafe}$ %
\end{theorem}
\begin{theorem}[$\cccf$ secure w.r.t. \gls*{ms}]\label{thm:cccf:rtp:ms}
  $\rtp{\cccf}{\mssafe}$ %
\end{theorem}

With both \Cref{thm:ccdce:rtp:ms,thm:cccf:rtp:ms} it follows from \Cref{corr:swappable} that the two passes can be interchanged arbitrarily:

\begin{theorem}[$\cccf\circ\ccdce$ and $\cccf\circ\ccdce$ are secure w.r.t. \gls*{ms}]\label{thm:cccfccdce:rtp:ms}
$\;$ 

  \begin{nscenter}
  \phantom{and~}\!\!$\rtp{\cccf\circ\ccdce}{\mssafe}$ 

  and~$\rtp{\ccdce\circ\cccf}{\mssafe}$ %
  \end{nscenter}
\end{theorem}

\subsection{Robust Strict Cryptographic Constant Time Preservation}\label{subsec:cs:scct}

This section defines a compiler $\ccscct$ from $\Lms$ to $\Lscct$ that robustly preserves \gls*{scct}. 
Given the fact that $\Lscct$ provides a \gls*{cct}-mode that can be turned on or off, the compiler inserts wrapper code for function calls and function bodies to ensure that execution in the component always happen in \gls*{cct}-mode.
This simple flag combines the effect of FaCT~\cite{cauligi2019fact} 

\vspace{-1em}
\[
\begin{array}{rcl}
  \ccscct(\irl{\lfunction{g}{x}{\expr}}) & = & \lfunction[\obj]{\obj{g}}{\obj{x}}{\obj{\lwrdoit{ON};}\ccscct(\irl{\expr})} \\
  \ccscct(\irl{\lcall{g}{\expr}}) & = & \lcall[\obj]{\obj{g}}{\ccscct(\irl{\expr})\obj{; \lwrdoit{ON}}} 
\end{array}
\]
The context can overwrite the flag and exit the mode, but upon invoking a function that is part of the component, the flag is set again.
Because of this, the corresponding cross-language trace relation $\sim^{\Lms}_{\Lscct}$, only relates events without secrets:%

\begin{center} 
  \typerulenolabel{xrel:scct:noleak}{}{\irl{\preevent;\comp}\sim^{\Lms}_{\Lscct}\obj{\preevent;\comp;\unlock}}
\end{center}

The compiler is secure w.r.t. \gls*{scct}: %

\begin{theorem}[$\ccscct$ secure w.r.t. \gls*{scct}]\label{thm:ccscct:rtp:scct}
  \small
  $\rtpsim{\ccscct}{\scctsafe}{\sim^{\Lms}_{\Lscct}}$ %
\end{theorem}

\subsection{Robust Speculative Safety Preservation}\label{subsec:cs:ss}

This section defines the final compilation pass $\ccspec$, which ensures that $\Lscct$ programs, which are assumed to be \gls*{ss}, stay \gls*{ss} at $\Lspec$-level. 
To do so, $\ccspec$ inserts a speculation barrier after branches, which is sufficient to harden the program against speculative leaks, since SPECTRE-PHT~\cite{kocher2019spectre} is the only speculative leak modeled in the semantics of $\Lspec$.

\vspace{-1em}
\[
\begin{array}{cl}
  &\ccspec{(\obj{\lifz{\expr[_0]}{\expr[_1]}{\expr[_2]}})} = 
  \\
  &\qquad\qquad \lifz[\ird]{\ccspec{\left(\obj{\expr[_0]}\right)}}{\ird{\lbarrier;}\ccspec{\left(\obj{\expr[_1]}\right)} \\&\qquad\qquad}{ \ird{\lbarrier;}\ccspec{\left(\obj{\expr[_2]}\right)}} 
\end{array}
\]

Clearly, the corresponding cross-language trace relation $\sim^{\Lscct}_{\Lspec}$ has only one non-trivial case: for branches, only relate them where speculation is blocked by a barrier:

\begin{center}
  \typerulenolabel{xrel:spec:if}{}{\obj{Branch\ n}\sim^{\Lscct}_{\Lspec}\ird{Branch\ n}\cdot\ird{Spec}\cdot\ird{Barrier}\cdot\ird{Rlb}}
\end{center}

The base-event relation above scales to full events by ensuring the missing annotations ($\obj{\comp;\securitytag{}}$ and $\ird{\comp;\securitytag{}}$) are the same.
With this relation, we prove that $\ccspec$ is secure with respect to \gls*{ss}.
\begin{theorem}[$\ccspec$ secure w.r.t. \gls*{ss}]\label{thm:ccspec:rtp:spec}
  \small$\rtpsim{\ccspec}{\sssafe}{\sim^{\Lscct}_{\Lspec}}$ %
\end{theorem}

\subsection{Robust Preservation of Memory Safety, Strict Cryptographic Constant Time, and Speculative Safety}

Finally, this subsection combines all previous results into one compilation chain to get that it preserves full \gls*{specms}.
Let $\ccspecms$ be the compiler that is the composition of $\cca$, $\ccb$, $\cccf$, $\ccdce$, $\ccscct$, and $\ccspec$. 
Let $\sim^{\Ltms}_{\Lspec}$ be the composition of $\sim^{\Ltms}_{\Lms}$, $\sim^{\Lms}_{\Lscct}$, and $\sim^{\Lscct}_{\Lspec}$.
Then, the following corollary holds.

\begin{corollary}[$\ccspecms$ secure w.r.t. \gls*{specms}]\label{thm:ccall:rtp:specms}
  $\;$ 

  \begin{nscenter}
    $\rtpsim{\cc{\Ltms}{\Lspec}}{\mssafe\cap\scctsafe\cap\sssafe}{\sim^{\Ltms}_{\Lspec}}$ %
  \end{nscenter}
\end{corollary}

As with \Cref{corr:ccab:rtp:ms}, it is important to ensure that the respective cross language trace relations are well-formed (\Cref{def:wfc:sig:tracerel}).
It is already known that $\sim^{\Ltms}_{\Lms}$ is well-formed with respect to $\mssafe$ (\Cref{lem:wf:ltmsltrg}).
Next in the chain is $\sim^{\Lms}_{\Lscct}$, which has to be well-formed w.r.t. $\scctsafe$.
This lemma holds, since a trace that was $\scctsafe$ is $\scctsafe$ even after applying $\sim^{\Lms}_{\Lscct}$: the relation enforces that $\Lscct$ traces related to $\Lms$ traces have no leaks of secrets whatsoever.

\begin{lemma}[$\sim^{\Lms}_{\Lscct}$ well-formed w.r.t. $\scctsafe$]\label{lem:wf:lsmslscct}
$\;$ 

  \begin{nscenter}
  $\wfcsig{\sim^{\Lms}_{\Lscct}}{\scctsafe}$
  \end{nscenter}
\end{lemma}

The last relation is $\sim^{\Lscct}_{\Lspec}$ which needs to be well-formed w.r.t. $\sssafe$.
Similarly to the previous relation, this holds, since $\sim^{\Lscct}_{\Lspec}$ only relates $\Lspec$ traces, which do not have speculative leaks, with $\Lscct$ traces.

\begin{lemma}[$\sim^{\Lscct}_{\Lspec}$ well-formed w.r.t. $\sssafe$]\label{lem:wf:lscctlspec}
  $\wfcsig{\sim^{\Lscct}_{\Lspec}}{\sssafe}$
\end{lemma}

\section{Formal Insights}\label{sec:formalities}

This section discusses how to connect each language-specific security property to the general security properties of \Cref{sec:compprop} (\Cref{subsec:formalities:maps}), and it demonstrates that the security property resulting from the universal image projection is faithful to the original property (\Cref{subsec:formalities:props}). 
Then, this section discusses why the order of compiler passes matters, and how does our framework help with identifying insecure compositions (\Cref{subsec:compatsecpasses}), and it gives additional technical insights on the secure compilation proofs (\Cref{subsec:seccompproofs}).

\subsection{From Language Traces to General Ones}\label{subsec:formalities:maps}

The previous theorems talk about preserving properties expressed in the trace model of the languages of each compiler.
However, these trace models are not the same trace model we used to specify the properties of \Cref{sec:compprop} (indicated with \ev{L}), which serves as the ``ground truth'' for the meaning of our properties.
To bridge this gap, the formal development requires specifying additional trace relations, from each of the language trace models to the \ev{L} one, that, for example, relate $\src{Get\ \loc\ n}$ and $\src{Set\ \loc\ n}$ to $\ev{Use\ \loc\ n}$ (and that induce a related universal image that we use in \Cref{subsec:formalities:props}).
One key insight of these relations is that they all omit context-made actions for two reasons: (1) contexts (which are universally quantified) can trivially invalidate any property and (2) we are interested in the component upholding the properties.

\subsection{Security Properties and Their Meaning}\label{subsec:formalities:props}
Each of the presented compilers use a cross-language trace relation, which is also used to translate the property from one language to the other one (via the existential or universal images).
While the meaning of projected properties does change with a translation, the change should not allow for a flawed compilation pipeline.
For example, we could be using a trace relation that translates a property in a language to a totally different one in another language.
To raise the trust into the translation of properties, \Cref{thm:prop-rel-corr} states (in a general fashion) that each security property is faithfully translated using the universal image according to the cross-language trace relation induced by the compiler.

\begin{theorem}[Properties Relation Correctness]\label{thm:prop-rel-corr}
  \begin{align*}
    \forall& \pi \in \{\tmssafe,\smssafe,\scctsafe,\sssafe\}, 
    \\
    \text{ for each }& \text{pair of languages } \src{L} \text{ and } \trg{L} \text{ used by the compilers},
    \\
    \text{ if }& 
    \sigma_{\sim^{\trg{L}}_{\ev{L}}}(\pi) \sim^{\trg{L}}_{\ev{L}} \pi
    \text{ and } 
    \sigma_{\sim^{\src{L}}_{\trg{L}}}(\sigma_{\sim^{\trg{L}}_{\ev{L}}}(\pi)) \sim^{\src{L}}_{\trg{L}} \sigma_{\sim^{\trg{L}}_{\ev{L}}}(\pi)
    \\
    \text{ then }& 
    \sigma_{\sim^{\src{L}}_{\trg{L}}}(\sigma_{\sim^{\trg{L}}_{\ev{L}}}(\pi)) \sim^{\src{L}}_{\ev{L}} \pi
  \end{align*}
\end{theorem}
The complexity of this theorem is that the relation in the conclusion cannot be obtained by composing the two relations in the premises.

We now informally argue why this theorem holds for the composition of all considered properties from \Cref{sec:compprop}.

\paragraph{$\sigma_{\sim^{\Ltms}_{\Lms}}\left(\tmssafe\cap\smssafe\cap\scctsafe\cap\sssafe\right)$}

For the four properties considered here, the trace models of $\Ltms$, $\Ltrg$, and $\Lms$ do not consider actions related to $\scctsafe$ and $\sssafe$, so these two properties are trivially translated correctly.
We now discuss the remaining $\tmssafe$ and $\smssafe$ in the form of their intersection $\mssafe$.

Recall that $\sim^{\Ltms}_{\Lms}$ is the composition of $\sim^{\Ltms}_{\Ltrg}$ and $\sim^{\Ltms}_{\Lms}$.
Since $\sim^{\Ltms}_{\Ltrg}$ is an equality, this relation trivially preserves the meaning of translated properties: related traces are identical!

Finally, let us consider a trace $\irl{\trace}\in\mssafe$ and understand what is that is related to via $\sim^{\Ltrg}_{\Lms}$.
{\em All} traces $\trg{\trace}$ with $\trg{\trace}\sim^{\Ltrg}_{\Lms}\irl{\trace}$ are identical to $\irl{\trace}$ except for get and set actions, which require for in-bound accesses (as stated in \Cref{subsec:cs:ms}): this clearly respect $\mssafe$.

\paragraph{$\sigma_{\sim^{\Lms}_{\Lscct}}\left(\tmssafe\cap\smssafe\cap\scctsafe\cap\sssafe\right)$}

As before, the trace models of $\Lscct$ and $\Lms$ cannot express $\sssafe$, so that is trivially translated correctly.
Also, the trace model of $\Lscct$ extends the one of $\Lms$ with respect to $\tmssafe$ and $\smssafe$ events, so the translation argument regarding those two properties is the same as before.
Thus, we need to reason about whether $\scctsafe$ is translated correctly.

By definition, $\sim^{\Lms}_{\Lscct}$ only relates $\obj{\unlock}$ events, which is also the same relation induced by $\sim_{\ctsafe}$ in \Cref{sec:msscct-rel}.
This ensures that composing the relations only relates $\unlock$ events, and thus the property is translated correctly.

\paragraph{$\sigma_{\sim^{\Lscct}_{\Lspec}}\left(\tmssafe\cap\smssafe\cap\scctsafe\cap\sssafe\right)$}
The trace model of $\Lspec$ extends the one of $\Lscct$ with respect to $\tmssafe$, $\smssafe$, and $\scctsafe$, so the translation argument for those three properties is the same as before.
Concerning $\sssafe$, $\sim^{\Lscct}_{\Lspec}$ relates $\Lspec$ speculation traces with $\Lscct$ branches with the $\lock_{\text{NONE}}$ tag, so the property is translated correctly, according to the relation defined in \Cref{sec:spec-ms-rel}.

\subsection{Compatibility of Secure Compiler Passes}\label{subsec:compatsecpasses}

Consider applying $\ccspec$ first and then $\ccb$ (albeit currently syntactically impossible).
In this case, $\ccb$ would insert new branches into the code that are not protected by a speculation barrier! 
This concern is reflected in the proofs that establish that the security class resulting of the composition of the trace relations is meaningful.
For the $\ccspec\cdot\ccb$ case, the class is $\sigma_{\sim_{\ird{\mathghost}}\bullet\sim_{\irl{ms}}}\left(\mssafe\cap\scctsafe\cap\sssafe\right)$.
Since $\ev{Use\ \loc\ n}\sim_{\irl{ms}}\irl{Branch\ n}\cdot\irl{Spec}\cdots$, where $\cdots$ does not contain a $\ev{{Barrier}}$ event, the resulting class is \emph{not} the original $\sssafe$ that is intended and it would break the corresponding \Cref{thm:prop-rel-corr}.

However, the composition is still technically possible and it is the job of the compiler engineers to ensure that the secure compilation pipeline happens in an order that ensures that the mapped security property is the intended one.

\subsection{Secure Compilation Proofs}\label{subsec:seccompproofs}

Our secure compilation proofs rely on backtranslations~\cite{abate2019jour,patrignani2021rsc}, which let one construct a source context starting from either target traces (aka trace-based backtranslations) or target contexts (aka context-based backtranslations).
These backtranslations also require setting up cross-language relations between various language elements such as expressions and program states, so we leave these details for the technical reports.
All backtranslations in the case-study are trace based except for those required by $\ccdce$ and $\cccf$, which are context-based (and they are an identity function). 

\section{Related Work\pages{2}}\label{sec:relwork}

This section discusses robust compilation, other secure compilation criteria and work related to the properties preserved in the case study.

\paragraph*{Secure Compilation as Robust Preservation}\label{subsec:relw:seccomprtp}

The robust preservation of properties as a compiler-level criterion has been analysed extensively~\cite{abate2019jour,patrignani2021rsc,abate2021extacc,patrignani2019survey} and thus we build on that framework.
No existing work is concerned with composing secure compilers, however, existing work~\cite{abate2021extacc} sketches composition of trace-relating compiler correctness in a similar way to what has been presented here.
The work relating robust preservation with universal composability~\cite{patrignani2022universal} is closest to what this paper presents.
The authors demonstrate a similar compositionality theorem to ours (\Cref{sec:sequential}) but use it in the context of protocols.
The work does not consider the generality to support different trace models or composition of compilers which robustly preserve different classes.

There is work on lifting exploits for single compilers to the whole chain~\cite{paykin2019weird}.
While that work considers {\em in}secure compilation and composition thereof in terms of exploits, the composition they are interested in allows to lift an exploit for one compiler pass to the whole compilation chain. 
Our framework takes the opposite direction and provides compositionality results for secure compilers.

\paragraph*{Other Secure Compilation Criteria}\label{subsec:relw:seccompcrit}

While this paper focuses on the robust preservation framework~\cite{abate2019jour}, other secure compilation criteria exist.
The survey on formal approaches to secure compilation~\cite{patrignani2019survey} discusses a broad spectrum already, while this section presents a very high-level overview.
Fully abstract compilation~\cite{abadi1999fullabstraction} states that a compiler should preserve and reflect observational equivalence between source and target programs.
Abate \emph{et al.}~\cite{abate2021faandrc} showed that fully abstract compilers robustly preserve program properties that are either trivial or meaningless.
As a mitigation for this, the authors presented a categorical approach based on maps of distributive laws~\cite{watanabe2002modl}, which they call many maps of distributive laws.
Maps of distributive laws have been investigated before as a possible secure compilation criterion~\cite{tsampas2020catsc}.
Other approaches are extensions of the compiler correctness criterion as discussed in other work~\cite{patterson2019next700} or the introduction of opaque observations~\cite{vu2021reconciling} to reconcile compiler optimisations with security.
Note that this work also presents secure compilers that are optimising, but contrary to the other~\cite{vu2021reconciling}, provides a formal account of these in the robust preservation framework.

\paragraph*{Memory Safety Mechanisms}\label{subsec:relw:msmechs}

Different mechanisms for enforcing memory safety exist that also consider the secure compilation domain, i.e., have an active attacker model.
For example, the ``pointers as capabilities'' principle represents pointers as machine-level capabilities~\cite{korashy2021capableptrs}, which behave in a similar fashion to capabilities by means of linear typing~\cite{morrisett2005L3}.
The approach of this paper also uses linear typing, but differs from $L^{3}$~\cite{morrisett2005L3} in the way that functions are not first-class.
Moreover, this paper considers an active attacker, while the work on $L^{3}$ only discusses whole programs and, thus, has no active attacker model.
The instrumentation to ensure memory safety that this paper presents is inspired by Softbounds~\cite{nagarakatte2009soft}.
That work inserts bounds-checks in front of pointer-dereferences and, for this to work, inserts meta-data information on pointer creation.
Softbounds also works in a more advanced setting with structured fields accesses and also introduces a table-lookup for pointers that are stored in memory.
This paper only considers arrays of primitive data, i.e., there are no pointers to pointers or structures.
Several other approaches to memory-safety exist in literature, specifically as compiler instrumentations~\cite{akritidis2009baggy,younan2010paricheck,jung2021pico,shankaranarayana2023tailcheck,dhumbumroong2020boundwarden,nam2019framer,zhou2023fatptrs}, hardware-extensions~\cite{kwon2013lowfat,saileshwar2022heapcheck,chen2023flexpointer,kim2023whistle}, or programming language extensions~\cite{elliott2018checkedc,li2022formalcheckedc,jim2002cyclone,elliott2015guilt,west2005cuckoo,weis2019fyr,benoit2019uniqueness}.
What differentiates this work from them is that this work uses known, compiler-based approaches to ensure memory-safety as a means to investigate secure compiler compositions.
This paper does not provide efficient memory-safety, but serves as a theoretical foundation for the secure compilation domain.

To extend the languages in this paper with a less restricted form of pointer arithmetic, the region colouring memory safety monitor presented in earlier work~\cite{michael2023mswasm} can be used.
The work presenting this monitor provides an approach for the robust preservation of memory safety compiling from C to WASM.
However, they do not discuss composition of secure compilers but rather investigate an instance of a secure compiler.

\paragraph*{Cryptographic Constant Time Mechanisms}\label{subsec:relw:cctmechs}

The approach to preserving cryptographic constant time in this paper is high-level, where a programming language exposes a way to switch the semantics to a data (operand) independent timing mode.
Since identifiers in $\Lscct$ are annotated with a secrecy tag, this approach is similar to others with information flow control.
For example, Vale~\cite{bond2017vale} uses Dafny to ensure constant-time assembly code, while Jasmin~\cite{almeida2017jasmin} makes use of the Coq proof assistant to reject non-constant-time programs.
CT-Wasm~\cite{watt2019ctwasm} enforces constant-timeness by means of a type system.
Different to the approach of this paper, these approaches necessitate that the programmer writes \gls*{cct} code.
An approach to allow programmers to write more high-level code is CryptOpt~\cite{kuepper2023cryptopt}, which generates efficient target-code by means of a randomised search.
This paper abstracts over concrete mitigation strategies and simply assumes that there is a flag to switch to a cryptographic-constant time execution mode.
This can be realised by employing the FaCT~\cite{cauligi2019fact} compiler, which translates common non-constant time code patterns to be constant-time, and the data (object) independent timing execution mode of modern processors.

\paragraph*{Speculation Safety Mechanisms}\label{subsec:relw:ssmechs}

This paper uses a taint-tracking mechanism inspired by existing work~\cite{guarnieri2018spectector,fabian2022automatic}.
These taints are used to express absence of any speculative leaks in \gls*{ss}~\cite{guarnieri2018spectector}. 
The semantics of $\Lspec$ hardcodes the kind of speculative leaks to just SPECTRE-PHT~\cite{kocher2019spectre}, but future work could use semantics composition~\cite{fabian2022automatic} to support more variants.
Note that our framework composes compilers and not semantics.

\section{Conclusion\pages{$\sfrac{1}{2}$}}\label{sec:concl}
This paper tackles the problem of understanding what kind of security properties a secure compiler preserves, when said compiler is the combination of compiler passes that preserve possibly different security properties.
The paper proves that composing secure compilers that preserve certain properties results in a secure compiler that preserves the composition of these properties.
Finally, this paper defines a multi-pass compiler and proves that it preserves \gls*{specms}.
Crucially, this paper derives the security of the multi-pass compiler from the composition of the security properties preserved by its individual passes, which include security-preserving as well as optimisation passes.

\newpage

\bibliographystyle{IEEEtranS}
\bibliography{main}

\appendix
\subsection{Existential Image}\label{subsec:extimg}

\begin{definition}[Existential Image]
\label{def:existential:img}\label{def:tau}
  \[
    \tau_\sim\left(\src{\pi}\right) := 
      \left\{ 
        \trg{\trace} \mid \exists \src{\trace}\ldotp \src{\trace}\sim\trg{\trace}, \text{ and } \src{\trace}\in\src{\pi} 
      \right\}
  \]
\end{definition}

\begin{definition}[Robust Preservation with $\tau_\sim$]\label{def:rtp:tau}
  {$\rtptau{\cc{\src{L}}{\trg{L}}}{\src{\class}}{\sim}$}
  $\isdef$
    {$\forall \src{\pi}\in\src{\class}, \src{p}\in\src{L},$} if {$\rsat{\src{\progvar}}{\src{\pi}}$}, then {$\rsat{\cc{\src{L}}{\trg{L}}\left(\src{p}\right)}{\tau_\sim\left(\src{\pi}\right)}$}.
\end{definition}

\begin{theorem}[Composition of Secure Compilers w.r.t. $\tau$]\label{thm:rtpsim:tau}
  $\;$ 

  If {$\rtptau{\cc{\src{L}}{\trg{L}}}{\src{\class[_{1}]}}{\sim_1}$}, {$\rtptau{\cc{\trg{L}}{\obj{L}}}{\tilde{\tau}_{\sim_1}\left(\src{\class[_2]}\right)}{\sim_2}$}, and {$\wfctau{\sim_1}{\src{\class[_2]}}$}, \\ then {$\rtptau{\cc{\src{L}}{\trg{L}}\circ\cc{\trg{L}}{\obj{L}}}{\src{\class[_{1}]}\cap\src{\class[_{2}]}}{\sim_1\bullet\sim_2}$}. \Coqed
\end{theorem}

\begin{corollary}[Swapping Secure Compiler Passes]\label{corr:swappable:tau}
  $\;$ 

  If {$\rtptau{\cc[_{1}]{\trg{L}}{\trg{L}}}{\src{\class[_{1}]}}{\sim_1}$ and $\rtptau{\cc[_{2}]{\trg{L}}{\trg{L}}}{\src{\class[_{2}]}}{\sim_2}$}, %
  {$\wfctau{\sim_1}{\src{\class[_2]}}$ and $\wfctau{\sim_2}{\src{\class[_1]}}$}, %
  and {$\tilde{\tau}_{\sim_1}\left(\src{\class[_2]}\right)=\src{\class[_2]}$ as well as $\tilde{\tau}_{\sim_2}\left(\src{\class[_1]}\right)=\src{\class[_1]}$},
  then {$\rtptau{\cc[_{1}]{\trg{L}}{\trg{L}}\circ\cc[_{2}]{\trg{L}}{\trg{L}}}{\src{\class[_{1}]}\cap\src{\class[_{2}]}}{\sim_1\circ\sim_2}$ and $\rtptau{\cc[_{2}]{\trg{L}}{\trg{L}}\circ\cc[_{1}]{\trg{L}}{\trg{L}}}{\src{\class[_{2}]}\cap\src{\class[_{1}]}}{\sim_2\circ\sim_1}$}. \Coqed
\end{corollary}

\subsection{Secure Upper and Lower Composition}\label{sec:other-compos}
Besides sequential composition, there are two other compositions, namely an {\em upper}, i.e., a compiler that takes multiple inputs and yields one output, and a {\em lower} composition, i.e., a compiler that takes one input and yields multiple outputs.
We {define the upper composition $\cc{\src{L}+\obj{L}}{\trg{L}}$} as follows:
Given a program \texttt{p}, its compiled counterpart is obtained by {plugging \texttt{p} into $\cc{\src{L}}{\trg{L}}$ if $\texttt{p}\in\src{L}$} or by {plugging \texttt{p} into $\cc{\obj{L}}{\trg{L}}$ if $\texttt{p}\in\obj{L}$}.
\begin{definition}[Upper Composition]
  \[
    \text{{$\cc{\src{L}+\obj{L}}{\trg{L}}$}}\isdef
  \lambda \texttt{p}\ldotp
  \left\{\mbox{\begin{tabular}{c}
    {{if $\texttt{p}\in\src{L}$, then $\cc{\src{L}}{\trg{L}}(\texttt{p})$}} \\
    \mbox{{if $\texttt{p}\in\obj{L}$, then $\cc{\obj{L}}{\trg{L}}(\texttt{p})$}} \\
  \end{tabular}}\right. 
  \]
\end{definition}

Examples of this are present in industry:
Consider the Java Virtual Machine bytecode $\trg{JVM BC}$, which is a popular target for programming language designers due to its high performance and relevance in industry.
Compilers for several programming languages have it as their target language, some popular instances are $\src{Java}$ and $\obj{Kotlin}$.
Technically speaking, they both compile to class files and $\obj{Kotlin}$ objects are considered to be the same as $\src{Java}$ objects at that point.
Both languages can be used at the same time in one project~\cite{androidstudio}.
A compiler that accepts both $\src{Java}$ and $\obj{Kotlin}$ code translating to the same target language or intermediate representation performs a kind of {\em upper} composition.
Now, the following theorem tells us what happens if these are secure:
Given {$\cc{\src{L}}{\trg{L}}$ robustly preserves $\class[_{1}]$} and {$\cc{\obj{L}}{\trg{L}}$ robustly preserves $\class[_{2}]$}, it follows that {their upper composition $\cc{\src{L}+\obj{L}}{\trg{L}}$ robustly preserves the intersection of classes $\class[_{1}]$ and $\class[_{2}]$}.

\begin{theorem}[Upper Composition of Secure Compilers]\label{thm:urtp}
  $\;$

  If {$\rtp{\cc{\src{L}}{\trg{L}}}{\class[_{1}]}$} and {$\rtp{\cc{\obj{L}}{\trg{L}}}{\class[_{2}]}$}, then {$\rtp{\cc{\src{L}+\obj{L}}{\trg{L}}}{\class[_{1}]\cap\class[_{2}]}$}. %
\end{theorem}

Dually, the {\em lower} composition is concerned about compilers that accept the same source but yield different target languages. %
{Define the lower composition $\cc{\src{L}}{\trg{L}+\obj{L}}$} as follows:
Given a program $\src{p}$, its compiled counterpart is obtained by {plugging $\src{p}$ into $\cc{\src{L}}{\trg{L}}$} or by {plugging \texttt{p} into $\cc{\src{L}}{\obj{L}}$}, respectively, {based on the internal decision}.
\begin{definition}[Lower Composition]
  \[
    \text{{$\cc{\src{L}}{\trg{L}+\obj{L}}$}}\isdef
  \lambda \src{p}, L\ldotp
  \left\{\mbox{\begin{tabular}{c}
    {{if $L=\trg{L}$, then} {$\cc{\src{L}}{\trg{L}}(\src{p})$}} \\
    \mbox{{if $L=\obj{L}$, then} {$\cc{\src{L}}{\obj{L}}(\src{p})$}} \\
  \end{tabular}}\right.\]
\end{definition}

 Consider two compilers both accepting $\src{LLVM IR}$~\cite{lattner2004llvm} and one of them emits $\trg{x86\_64}$, while the other emits $\obj{ARMv8}$.
 It is intuitive that they are in some sense composed in the LLVM framework, but the decision of when to use one over the other is inherently {\em internal} to the formalisation effort of this kind of composition.
 For example, the user of this compiler provides an explicit flag that instructs to emit $\trg{x86\_64}$ or the framework itself detects the target platform via heuristics, such as supported instructions.

 The following theorem demonstrates what happens if the involved compilers are secure:
Given {$\cc{\src{L}}{\trg{L}}$ robustly preserves $\class[_{1}]$} and {$\cc{\src{L}}{\obj{L}}$ robustly preserves $\class[_{2}]$}, it follows that {their lower composition $\cc{\src{L}}{\trg{L}+\obj{L}}$ robustly preserves the intersection of classes $\class[_{1}]$ and $\class[_{2}]$}.

\begin{theorem}[Lower Composition of Secure Compilers]\label{thm:lrtp}
  $\;$ 

  If {$\rtp{\cc{\src{L}}{\trg{L}}}{\class[_{1}]}$} and {$\rtp{\cc{\obj{L}}{\trg{L}}}{\class[_{2}]}$}, then {$\rtp{\cc{\src{L}}{\trg{L}+\obj{L}}}{\class[_{1}]\cap\class[_{2}]}$}. %
\end{theorem}

Either way, the theoretical results suggest that it is possible to always find a ``most-general'', secure compiler, given two secure compilers, that robustly preserves the least-upper bound of the classes involved in their compilation process.

\end{document}